\documentclass[aps,twocolumn]{revtex4}

\usepackage{enumerate}
\usepackage{amsmath}
\usepackage{amssymb}
\usepackage{amsthm}
\usepackage{bbm}
\usepackage{braket}
\usepackage{graphicx}

\newcommand{\lk}{\left(}
\newcommand{\rk}{\right)}

\newcommand{\Id}{\mathbbm{1}}

\newcommand{\GHZ}{\ensuremath{\ket{\mathrm{GHZ}}} }

\newcommand{\QED}{\hfill\ensuremath{\Box}\vspace{.6cm}}

\def\clap#1{\hbox to 0pt{\hss#1\hss}}

\newtheorem{theorem}{Theorem}
\newtheorem{lemma}{Lemma}
\newtheorem{definition}{Definition}
\newtheorem{corollary}{Corollary}
\newtheorem{remark}{Remark}
\newtheorem{example}{Example}

\begin{document}

\title{Bounds on probability of transformations between multi-partite
pure states}

\author{Wei \surname{Cui}}
\email[]{cuiwei@physics.utoronto.ca}
\affiliation{Center for Quantum Information and Quantum Control (CQIQC),
Department of Physics and Department of Electrical \& Computer Engineering,
University of Toronto, Toronto, Ontario, M5S 3G4, Canada}

\author{Wolfram \surname{Helwig}}
\email[]{whelwig@physics.utoronto.ca}
\affiliation{Center for Quantum Information and Quantum Control (CQIQC),
Department of Physics and Department of Electrical \& Computer Engineering,
University of Toronto, Toronto, Ontario, M5S 3G4, Canada}

\author{Hoi-Kwong \surname{Lo}}
\email[]{hklo@comm.utoronto.ca}
\affiliation{Center for Quantum Information and Quantum Control (CQIQC),
Department of Physics and Department of Electrical \& Computer Engineering,
University of Toronto, Toronto, Ontario, M5S 3G4, Canada}

\begin{abstract}
For a tripartite pure state of three qubits, it is well known that
there are two inequivalent classes of genuine tripartite entanglement,
namely the GHZ-class and the W-class. Any two states within the same
class can be transformed into each other with stochastic local
operations and classical communication (SLOCC) with a non-zero
probability. The optimal conversion probability, however, is only
known for special cases. Here, we derive new lower and upper bounds
for the optimal probability of transformation from a GHZ-state to
other states of the GHZ-class. A key idea in the derivation of the upper
bounds is to consider the action of the LOCC protocol on a different input
state, namely $1/\sqrt{2} [\ket{000} - \ket{111}]$, and demand that
the probability of an outcome remains bounded by 1. We also find an upper bound for more general cases by using the constraints of the so-called interference term and 3-tangle.
Moreover, we generalize some of our results to the case where each
party holds a higher-dimensional system. In particular, we found that
the GHZ state generalized to three qutrits, i.e., $\ket{\mathrm{GHZ}_3} =
1/\sqrt{3} [ \ket{000} + \ket{111} + \ket{222} ] $, shared
among three parties can be transformed to {\it any} tripartite 3-qubit pure state
with probability 1 via LOCC. Some of our results can also be generalized to the
case of a multipartite state shared by more than three parties.
\end{abstract}

\maketitle
%\tableofcontents

\section{Introduction}
Entanglement is the most peculiar feature that distinguishes quantum
physics from classical physics and lies at the heart of quantum
information theory. Thus it is important to get a good understanding
of entanglement properties of quantum states. These properties are
well understood for bipartite pure states. In the standard distant
laboratory paradigm,  suppose two distant parties, Alice and Bob,
shared a bipartite entangled state. They may apply local operations
and classical communications (LOCC) to convert it into another
partite state. Bennett et al \cite{Bennett1996}  has answered the
question for the rate of LOCC transformation between bipartite pure
states. It is quantified by the von Neumman entropy of a reduced
density matrix. For the single-copy case, the optimal conversion
probabilities are known for any pure state transformation
\cite{Lo2001, Nielsen1999, Vidal1999}. For an LOCC transformation
protocol, if it can succeed with probability 1, we call it
deterministic, if it can only succeed with a nonzero probability
smaller than 1, we call it stochastic, or SLOCC (Stochastic Local
Operators and Classical Communications). For mixed states, the
question of what the optimal rate of transformations is between them
is still largely open.

For multipartite states, however, the problem is much more complicated.
There exist different types of entanglement and therefore the
transformations are rather involved. For the case of tripartite pure three
qubit states, a characterization into six different entanglement classes,
of which two contain true tripartite entanglement, exists \cite{Dur2000}. One is the GHZ class state, which is defined as

\begin{equation}
\ket{\phi_{GHZ}} = \sqrt{K}(c_{\delta}\ket{0}\ket{0}\ket{0} +
s_{\delta}e^{i\varphi}\ket{\varphi_A}\ket{\varphi_B}\ket{\varphi_C})
\end{equation}
where
\begin{eqnarray}
\ket{\varphi_A} =& c_{\alpha}\ket{0} + s_{\alpha}\ket{1}\\
\ket{\varphi_B} =& c_{\beta}\ket{0} + s_{\beta}\ket{1}\\
\ket{\varphi_C} =& c_{\gamma}\ket{0} + s_{\gamma}\ket{1}
\end{eqnarray}

and K=$(1 +
2c_{\delta}s_{\delta}c_{\alpha}c_{\beta}c_{\gamma}c_{\phi})^{-1} \in [\frac{1}{2}, \infty)$,
$c_{\delta} = \cos\delta, s_{\delta} = \sin\delta$, the same for $\alpha,
\beta, \gamma, \phi$. The range for the parameters are $\delta \in (0, \frac{\pi}{4}]$, $\alpha, \beta, \gamma \in (0, \frac{\pi}{2}]$ and $\varphi \in [0, 2\pi)$.

Another is W class state, which is defined as a state that is unitarily equivalent to

\begin{equation}
\ket{\phi} = (\sqrt {c}\ket{0} + \sqrt{d}\ket{1}) \ket{00} + \ket{0}(\sqrt{a}\ket{01} + \sqrt{b}\ket{10})
\end{equation}
with $c+d+a+b =1$.

A transformation between any two states of the same class is always possible
with non-zero probability. However, here comes the key point. The optimal conversion between the states within the same class of genuine tripartite entangled states is \emph{not} known. Incidentally, a
similar characterization into nine different classes exists for four qubits
\cite{Verstraete2002}. In 2000, the optimal rate of distillation of a GHZ state from any GHZ-class state was found \cite{Ac'in2000a}. Very recently, a necessary and sufficient
condition for deterministically (i.e., with probability 1) transforming
multipartite qubit states with Schmidt rank 2 \cite{Eisert2001} have been given \cite{Turgut2009}.

In this paper, we present new upper and lower bounds for multipartite
entanglement transformations. In particular, we focus on transformations among states with the
same Schmidt rank \cite{Eisert2001}. We put an
emphasis on the transformation from a GHZ state to a GHZ-class state.

But our upper bound can also be generalized to general transformations from one GHZ class state to another. And some of the results
are derived for the more general case of higher dimensions and more
parties. Especially, we find that all tripartite pure three qubit states can be transformed from 3-term GHZ
state
$\frac{1}{\sqrt{3}}(\ket{000}+\ket{111}+\ket{222})$ with probability
1. This is a new result. Moreover, some of the general theorems for deterministic transformation are also derived.

The paper is structured as follows. In Section~\ref{sec:upperbound1}, we
derive upper bounds for the transformation of the GHZ-state to any other
state in the GHZ-class. The upper bounds are only non-trivial for a subclass
of the GHZ-class. Thus Section~\ref{sec:failurebranch} and
\ref{sec:upperbound2} use a different
approach that results in upper bounds for a wider class of states. More specific, for any GHZ class state which does not have a known way to be transformed from GHZ state with probability 1, we can find a nontrivial upper bound for the probability of this transformation. And our upper bound can also be effective for the transformation from a GHZ class state to a large class of other GHZ class states. Lower
bounds for the transformation of higher dimensional GHZ-states distributed
among three or more parties to states with the same Schmidt rank are given
in Section~\ref{sec:lowerbound}.

\section{Upper Bound for the Conversion from GHZ state to a GHZ class state}
\label{sec:upperbound1}

In this section, we derive an upper bound for the conversion of the
GHZ-state to any other state of the GHZ-class via LOCC. This upper bound
will be nontrivial (i.e., smaller than 1) for $\varphi \in (\frac{1}{2} \pi,
\frac{3}{2} \pi)$. The transformation under consideration is given by
\begin{eqnarray}
\label{eq:GHZclass}
&\GHZ = \frac{1}{\sqrt{2}} (\ket{000} + \ket{111}) \notag\\
\stackrel{\textrm{LOCC}} {\longrightarrow} \ket\Psi &= \sqrt{K}(c_{\delta}\ket{0}\ket{0}\ket{0} + s_{\delta}e^{i\varphi}\ket{\varphi_A}\ket{\varphi_B}\ket{\varphi_C}),
\end{eqnarray}
with the parameters defined in introduction.

The LOCC operation
is represented by Kraus operators $\{O_i = A_i \otimes B_i \otimes
C_i\}$. In the following we will refer to different Kraus operators of the
LOCC protocol as different \emph{branches}. Furthermore, a branch $O_i \GHZ
= \ket\Phi$ is called a
\emph{success branch} if $\ket\Phi \propto \ket\Psi$, and a
\emph{failure branch} if there exists no LOCC-operation that can transform
$\ket\Phi$ into $\ket\Psi$ with a non-zero probability, if a branch is neither success nor failure, we call it an \emph{undecided branch}. An optimal
protocol only consists of success and failure branches.

\begin{figure}[htp]
\centering
\includegraphics[width=0.5\textwidth, height=0.33\textwidth]{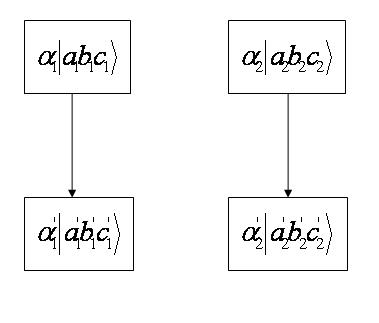}
\caption{mapping type 1}
\label{fig:upperbound11.jpg}
\end{figure}

\begin{figure}[htp]
\centering
\includegraphics[width=0.5\textwidth, height=0.33\textwidth]{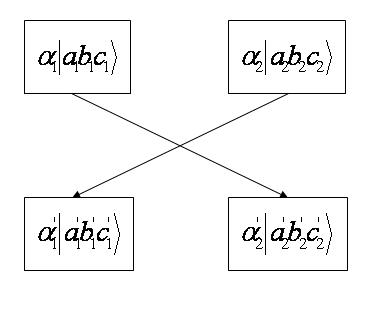}
\caption{mapping type 2}
\label{fig:upperbound12.jpg}
\end{figure}

For the following analysis we first recall two known results
\cite{Dur2000, Ac'in2000a}
\begin{lemma}
    \label{lem:prod_unique}
    For a GHZ-class state $\ket\Psi$ we have:
    \begin{enumerate}[a)]
      \item The Schmidt rank of $\ket\Psi$ is 2 \cite{Dur2000}. This means that the
            minimum number of product states necessary to write $\ket\Psi$ as a
            superposition of them is 2:
            \begin{equation}
                \label{eq:prod_state}
              \ket\Psi = \sum_{i=1}^{2} \alpha_i \ket{a_i b_i c_i},
            \end{equation}
            with $\alpha_i \in (0,1)$ and $\braket{a_i b_i c_i |a_i b_i c_i } =1$.
        \item This product state decomposition, i.e., the set $\{(\alpha_1,
            \ket{a_1 b_1 c_1}), (\alpha_2, \ket{a_2 b_2 c_2})\}$ is unique
            \cite{Ac'in2000}.
    \end{enumerate}
%
%
%   Any state $\ket{\Psi}$ of the GHZ-class can be uniquely written as a
% superposition of two, in general nonorthogonal, product states
% \begin{equation}
%   \label{eq:prod_state}
%   \ket{\Psi} = \alpha\ket{abc} + \beta\ket{a'b'c'}.
% \end{equation}
% with $\alpha, \beta \in (0,1)$ and $\braket{abc|abc} =
% \braket{a'b'c'|a'b'c'} = 1$.
\end{lemma}
This result leads to
\begin{lemma}
 \label{lem:locc_terms}
 For a successful LOCC operation within the GHZ-class,
 \begin{eqnarray}
  \ket{\Psi} &= \alpha_1 \ket{a_1b_1c_1} + \alpha_2\ket{a_2b_2 c_2}\notag\\
  \stackrel{\textrm{LOCC}}{\longrightarrow}
  \ket{\Psi'} &= \alpha'_1 \ket{a'_1b'_1c'_1} + \alpha'_2\ket{a'_2b'_2 c'_2},
 \end{eqnarray}
  described by the operator $O_1$, we must either have the mapping
 \begin{align}
    O_1 \ket{a_1b_1c_1} &= o_1 \frac{\alpha'_1}{\alpha_1} \ket{a'_1b'_1c'_1}\\
    O_1 \ket{a_2b_2c_2} &= o_1 \frac{\alpha'_2}{\alpha_2} \ket{a'_2b'_2c'_2}
 \end{align}
 or
 \begin{align}
    O_1 \ket{a_1b_1c_1} &= o_1 \frac{\alpha'_2}{\alpha_1} \ket{a'_2b'_2c'_2}\\
    O_1 \ket{a_2b_2c_2} &= o_1 \frac{\alpha'_1}{\alpha_2} \ket{a'_1b'_1c'_1}
 \end{align}
 with some proportionality constant $o_1$, which can be chosen to be real. See Figure~\ref{fig:upperbound11.jpg}, Figure~\ref{fig:upperbound12.jpg}.
\end{lemma}

Proof: Since a LOCC Kraus operator is always of the form $O_1 = A_1
\otimes B_1 \otimes C_1$, a product state is always
transformed into a product state. With that observation and
the fact that the two-term product decomposition of a tripartite
GHZ-class state is unique (Lemma~\ref{lem:prod_unique}), Lemma 2
follows. \QED

\begin{theorem}
\label{theo:upperbound}
    An upper bound for the conversion probability for
    \begin{eqnarray}
        \label{eq:theo1}
        &\GHZ = \frac{1}{\sqrt{2}} (\ket{000} + \ket{111})\notag\\
        \rightarrow
        \ket\Psi &= \sqrt{K}(c_{\delta}\ket{000} + s_{\delta}e^{i\varphi}\ket{\varphi_A\varphi_B\varphi_C}),
    \end{eqnarray}
where the parameters are defined in Equation~\eqref{eq:GHZclass},
is given by
    \begin{equation}
        p \leq \min\left\{1,
            \frac{1 + 2 c_{\delta}s_{\delta}c_{\alpha}c_{\beta}c_{\gamma}
            c_{\varphi}}{1 - 2 c_{\delta}s_{\delta}c_{\alpha}c_{\beta}c_{\gamma}c_{\varphi}}
            \right\}
    \end{equation}
\end{theorem}
Idea of the Proof: From Lemma \ref{lem:locc_terms} we know that, for a success branch, each product state (in the Schmidt term) of the input states have to be mapped to a product state of the
output state. This allows us to infer how the same LOCC protocol acts on the
phase flipped GHZ state, i.e., $\frac{1}{\sqrt{2}} (\ket{000} - \ket{111})$. From the requirement
that the sum of the probabilities for the output states have to sum to 1 for
this transformation, we can derive a bound for the parameters arising in the
original transformation. This gives a bound on the successful
transformation probability.

Proof: Consider the optimal LOCC strategy, given by the Kraus operators
$O_i = A_i \otimes B_i \otimes C_i$. According to
Lemma~\ref{lem:locc_terms}, there are two possibilities to have a
successful branch. They are
\begin{align}
  O_i \ket{000} &= o_i c_{\delta} \ket{000}\\
    O_i \ket{111} &= o_i e^{i\varphi}s_{\delta} \ket{\varphi_A\varphi_B\varphi_C}
\end{align}
for $i = 1, \ldots, n_1$, and
\begin{align}
  O_i \ket{000} &= o_i e^{i\varphi}s_{\delta} \ket{\varphi_A\varphi_B\varphi_C}\\
    O_i \ket{111} &= o_i c_{\delta} \ket{000}
\end{align}
for $i = n_1+1, \ldots, n_1+n_2$. Both cases give the desired transformation
\begin{equation}
  O_i \GHZ = \frac{1}{\sqrt{2}} o_i (c_{\delta} \ket{000} +
            e^{i\varphi}s_{\delta} \ket{\varphi_A\varphi_B\varphi_C})
        = \frac{o_i}{\sqrt{2K}} \ket\Psi
\end{equation}
for $i = 1, \ldots, n_1+n_2$. The successful conversion probability is then
given by
\begin{equation}
    \label{eq:prob}
  p = \frac{1}{2K} \sum_{i=1}^{n_1+n_2} o_i^2.
\end{equation}
To get an upper bound for $\sum_{i=1}^{n_1+n_2} o_i^2$, we consider how
\begin{equation}
  \frac{1}{\sqrt{2}} (\ket{000} - \ket{111})
\end{equation}
behaves when put through the Kraus Operator $O_i$. We see that
\begin{eqnarray}
    &O_i \frac{1}{\sqrt{2}} (\ket{000} - \ket{111})\notag\\
  = &\frac{1}{\sqrt{2}} o_i (c_{\delta} \ket{000} - e^{i\varphi}s_{\delta} \ket{\varphi_A\varphi_B\varphi_C}) = \frac{o_i}{\sqrt{2K'}} \ket{\Psi'}
\end{eqnarray}
with
\begin{equation}
  \ket{\Psi'} = \sqrt{K'} (c_{\delta} \ket{000} - e^{i\varphi}s_{\delta} \ket{\varphi_A\varphi_B\varphi_C}),\\
\end{equation}
where $K' = \frac{1}{1 - 2 c_{\delta}s_{\delta}c_{\alpha}c_{\beta}c_{\gamma}c_{\varphi}}$,
for $i = 1, \ldots, n_1+n_2$ (up to an overall minus sign for $i = n_1+1,
\ldots n_1+n_2$). Thus the conversion probability for this process is given
by $\frac{1}{2K'} \sum_{i=1}^{n_1+n_2} o_i^2$. Being a probability, this has
to be bounded by 1,
giving $\sum_{i=1}^{n_1+n_2} o_i^2 \leq 2 K'$. This together with
Equation~\eqref{eq:prob} gives the upper bound
\begin{equation}
  p \leq \frac{K'}{K} =
        \frac{1 + 2 c_{\delta}s_{\delta}c_{\alpha}c_{\beta}c_{\gamma}
            c_{\varphi}}{
        1 - 2 c_{\delta}s_{\delta}c_{\alpha}c_{\beta}c_{\gamma}
            c_{\varphi}}
\end{equation}
for the process described by Equation~\eqref{eq:theo1}. \QED

\emph{Special Case:} Regarding the special case, where we have $\ket{\varphi_A} =
\ket{\varphi_B} = \ket{\varphi_C}$, $c_{\alpha} = c_{\beta} = c_{\gamma} = \lambda_a$, $\varphi=0$, and $c_{\delta} = s_{\delta} = \frac{1}{\sqrt{2}}$,
i.e.,
\begin{equation}
  \ket\Psi = \frac{1}{\sqrt{2}\sqrt{1-\lambda_a^3}}
        (\ket{000} - \ket{aaa}),
\end{equation}
we get
\begin{equation}
  p \leq \frac{1 - \lambda_a^3}{1+\lambda_a^3}.
\end{equation}

Theorem~\ref{theo:upperbound} gives a non-trivial upper bound for the
transformation from the GHZ-state to a GHZ-class state for all values of
$\varphi$ with $\cos\varphi < 0$, i.e., $\phi \in ( \frac{\pi}{2} , \frac{3\pi}{2} )$. This nicely shows, that unlike in the
bipartite case, where the maximally entangled EPR-state can be
tranformed into any other pure two qubit state with probability one, the
GHZ-state, which exhibits maximal genuine tripartite entanglement as it
maximizes the 3-tangle \cite{Coffman2000} and tracing out one qubit results in a totally mixed
state, cannot be transformed to all other states in the same class with
probability one.

\section{Failure Branch}
\label{sec:failurebranch}

Recall in the last section that Eq. (22) gives a trivial bound for
the case $\phi \in ( \frac{\pi}{2} , \frac{3\pi}{2} )$.
Here, we will derive a useful bound for a larger class of states: we find a upper bound nontrivial for all the cases except $\phi = \frac{\pi}{2} or \frac{3\pi}{2}$ and $\braket{000|\varphi_A\varphi_B\varphi_C} = 0$. In fact, it was shown that these two kinds of transformations can succeed with probability 1 \cite{Turgut2009}.
Our proof has two important ingredients, namely, the conservation of
a new quantity defined as "interference term" under positive operator valued measures (POVMs) and that
the three tangle is an entanglement monotone, which we will discuss
in detail in the following.

The idea of our discussion is that, firstly, recall our definition of "failure branch" as one can not be successful with any nonzero probability, we will prove the weight summation of the so-called interference terms and normalizations of all the branches
in an LOCC protocol should be constant during the transformation, which is included in section 3. After that, we
find that three tangle is bounded for a fixed interference term
which will be defined in this section. Then, we try to see the whole
process from the weak measurement aspect, which divides the whole
process into many infinitesimal steps and each step changes the state
very little. That is to say, the state is changing continuously.
Then we stop in the middle and investigate whether there will be a
new upper bound. Surprisingly we find there are some new upper
bounds and these upper bounds will still be effective in the
following steps, even when we reach the end. So it can be used to
derive a new upper bound for the supremum success probability of the
whole LOCC protocol. Detailed discussion will be showed in section
4.

\begin{theorem}
For the transformation from GHZ to GHZ-class state $\ket\phi$, failure
branches should end with a state with at least one parties' reduced matrix
with rank 1.
\end{theorem}

Proof: Suppose we would like to get a GHZ-class state $\ket{\phi} =
\sqrt{K}(c_{\delta}\ket{0}\ket{0}\ket{0} +
s_{\delta}e^{i\varphi}\ket{\varphi_A}\ket{\varphi_B}\ket{\varphi_C})$, where
$\ket{0_A}$ is linearly independent of $\ket{\varphi_A}$, the same for B and
C. If there is a state whose reduced density matrices are all of full rank,
$\ket{\phi} = \sqrt{K'}(c'_{\delta}\ket{0}\ket{0}\ket{0} +
s'_{\delta}e^{i\varphi'}\ket{\varphi'_A}\ket{\varphi'_B}\ket{\varphi'_C})$,
where $\ket{0_A}$ is linearly independent of $\ket{\varphi'_A}$, the same
for B and C. Then it is easy to see, the equation \begin{eqnarray}
O_A \ket{0} &= \ket{0}\\
O_A \ket{\varphi'_A} &= \ket{\varphi_A}
\end{eqnarray}
always has a non-trivial solution, the same for B and C. That means
we can always transform this state into $\ket\phi$ with nonzero
probability. \QED

\subsection{Conservation of interference term}

To go further, we want to use the following property of the LOCC Kraus
operators. For a complete set of Kraus operators ${O_i=
A_i\otimes B_i\otimes C_i}$, we have $\sum {O^+_iO_i}=1$.

Suppose that a Kraus operator O satisfies
\begin{equation}
O\ket{000}= \alpha\ket{a_1b_1c_1}
\end{equation}
\begin{equation}
O\ket{111}= \beta\ket{a_2b_2c_2}
\end{equation}
with $\braket{a_1b_1c_1|a_1b_1c_1} = \braket{a_2b_2c_2|a_2b_2c_2} =1$.

Then it can transform
$\ket{GHZ}=\frac{1}{\sqrt{2}}(\ket{000}+\ket{111})$ into $\ket\psi=
\frac{1}{\sqrt{2p}}(\alpha\ket{a_1b_1c_1}+\beta\ket{a_2b_2c_2}$,
where $\frac{1}{\sqrt{2p}}$ is the normalization factor and one can check p is exactly the probability of getting $\ket\psi$. From here we
define interference term and normalization in the
following:

\begin{definition}
For a normalized GHZ-class state $\ket\gamma$ where
$\braket{\gamma|\gamma} = 1$, written in the form
$\ket\gamma=
\frac{1}{\sqrt{2}}(\alpha\ket{a_1b_1c_1}+\beta\ket{a_2b_2c_2})$,
suppose $\braket{a_1b_1c_1|a_2b_2c_2}= k$, then we
call the real part of $\alpha^* \beta k $ the interference term
$I$ of
$\ket\gamma$.
\end{definition}

It is easy to see if an operator O transforms $\ket{GHZ}$ to a state
$\ket\psi$, the interference term of $\ket\psi$ is in fact the real part of
$\frac{<000|O^+O|111>}{p}$, where p is the probability of the branch
corresponding to operator O.

\begin{remark}
In fact, one can find $I=  1 - \frac{1}{2}(|\alpha|^2 +
|\beta|^2)$.
\end{remark}
\begin{remark}
Note also that $-\infty < I \leq 1$. In other words, it can be unbounded
below.
This fact will become important in our discussion in Section 4.
\end{remark}
\begin{remark}
Notice that a failure branch gives a state that is $\mathbf{outside}$ the GHZ class.
For such a state, the actual value of interference term depends not only
on the state itself, but also on the particular Kraus operator, $O_i$, and
the initial state, $\phi_i$, used to reach the state. So, when we talk about the interference term of failure branches of an SLOCC transformation, we need to be careful: We are not talking about the interference term of the state given by the failure branches, but the interference term determined by the whole transformation protocol.
\end{remark}

\begin{theorem}
For a complete set of LOCCs which transforms GHZ state to other
states, in which the operators are $\{O_i\}$, the weighted sum of the
interference terms in all the branches should be zero.
\begin{equation}
0 = \sum p(O_i\ket{GHZ})I(O_i\ket{GHZ})
\end{equation}
where $p(O_i\ket{GHZ})$ is the probability of branch corresponding to the Kraus operator $O_i$, and $I(O_i\ket{GHZ})$ denotes the interference term I for a state $O_i\ket{GHZ}$.
\end{theorem}
Proof: Suppose the corresponding complete set of Kraus operators consists of  ${O_i= A_i\otimes B_i\otimes C_i}$. Then we have $\sum
{O^+_iO_i}= \Id$. So, we should have
\begin{eqnarray}
\label{eq: interference}
0=& <000|111> = <000|\Id|111>\notag\\
 =& <000|\sum {O^+_iO_i}|111>\notag\\
 =& \sum{<000|O^+_iO_i|111>}\notag\\
 =& \sum{p(O_i\ket{GHZ})\frac{<000|O^+_iO_i|111>}{p(O_i\ket{GHZ})}}
\end{eqnarray}

From the definition of interference term $I$ we know the real part of the right side of
Equation~\eqref{eq: interference} is exactly the weighted sum of $I$ of each branch. As
the right side of Equation~\eqref{eq: interference} is equal to zero, its real part
should also be zero, which means for a transformation from
$\ket{GHZ}$ to other states, the average value of the interference
terms of all the states we get in each branch should be zero. We
call this the \emph{conservation of interference term}. \QED

\subsection{Conservation of normalization}
\begin{definition}
For a two-term tripartite state $\ket\gamma$, written in the form
$\ket\gamma=
\frac{1}{\sqrt{2}}(\alpha\ket{a_1b_1c_1}+\beta\ket{a_2b_2c_2}$,  then we
call $\frac{1}{2}(|\alpha|^2+|\beta|^2)$ the normalization of
$\ket\gamma$.
\end{definition}

Easy to see if an operator O transforms $\ket{GHZ}$ to the state $\ket\psi$,
the normalization of $\ket\psi$ is in fact
$\frac{<000|O^+O|000>+<111|O^+O|111>}{2p}$, where p is the probability. And
because O is a positive operator, normalization should be always no less
than zero.

Suppose the corresponding complete set of Kraus operators consists of $\{O_i =$
 $A_i \otimes B_i \otimes C_i\}$. Then we have
$\sum {O^+_iO_i}= \Id$. So we should have

\begin{eqnarray}
\label{eq: normalization}
1=& <GHZ|GHZ>\notag\\
 =& \frac{1}{2}(<000|+<111|)(|000>+|111>)\notag\\
 =& \frac{1}{2}(\braket{000|000}+\braket{111|111})\notag\\
 =& \frac{1}{2} (\sum{\braket{000|O^+_iO_i|000}} +
\sum{\braket{111|O^+_iO_i|111}})\notag\\
 =& \sum{p(O_i\ket{GHZ})\frac{<000|O^+_iO_i|000> + <111|O^+_iO_i|111>}{2p(O_i\ket{GHZ})}}
\end{eqnarray}

From the definition of normalization we know it is exactly the
weighed sum of the normalization of each branch. That is to say,
for a transformation from $\ket{GHZ}$ to other states, the average value of
the normalization of all the states we get in each branch should be 1. And
recall that normalization can be no less than zero. So each term in the
summation
should be no larger than 1, which means for each branch, the product of its
probability and the normalization of the state it gets should be no larger
than 1.

In fact, the conservation of normalization can be derived from
conservation of interference term. However, conservation of the normalization
also gives the following. For each branch, the product of its
probability and the normalization of the state it gets should be no
larger than 1. The fact is also useful in determining the upper bound of
transformation probability.

The basic idea is that, if we know the state we want and the state
failure branch gives, equations \eqref{eq: interference} and \eqref{eq: normalization} combined with the fact that the summation of probability should be one can give us some implication about the supremum success probability. For example, we can have the following theorem:

\begin{theorem}
\label{theorem:failurebound} For a transformation protocol from GHZ
state to a GHZ-class state $\ket{\phi}$ whose interference term is
x, which is positive (negative), if there exists a y (y $>$ 0), such
that, the interference term of all the failure branches are larger
than -y (smaller than y), we have an upper bound for its successful
probability $p_s$ in the following:

if $x > 0$:

\begin{equation}
p_s \leq p^U(-y) = \frac{y}{x+y}.
\end{equation}
if $x < 0$:
\begin{equation}
p_s \leq p^U(y) = - \frac{y}{x-y}.
\end{equation}

See
\end{theorem}
\begin{figure}[htp]
\centering
\includegraphics[width=0.5\textwidth, height=0.33\textwidth]{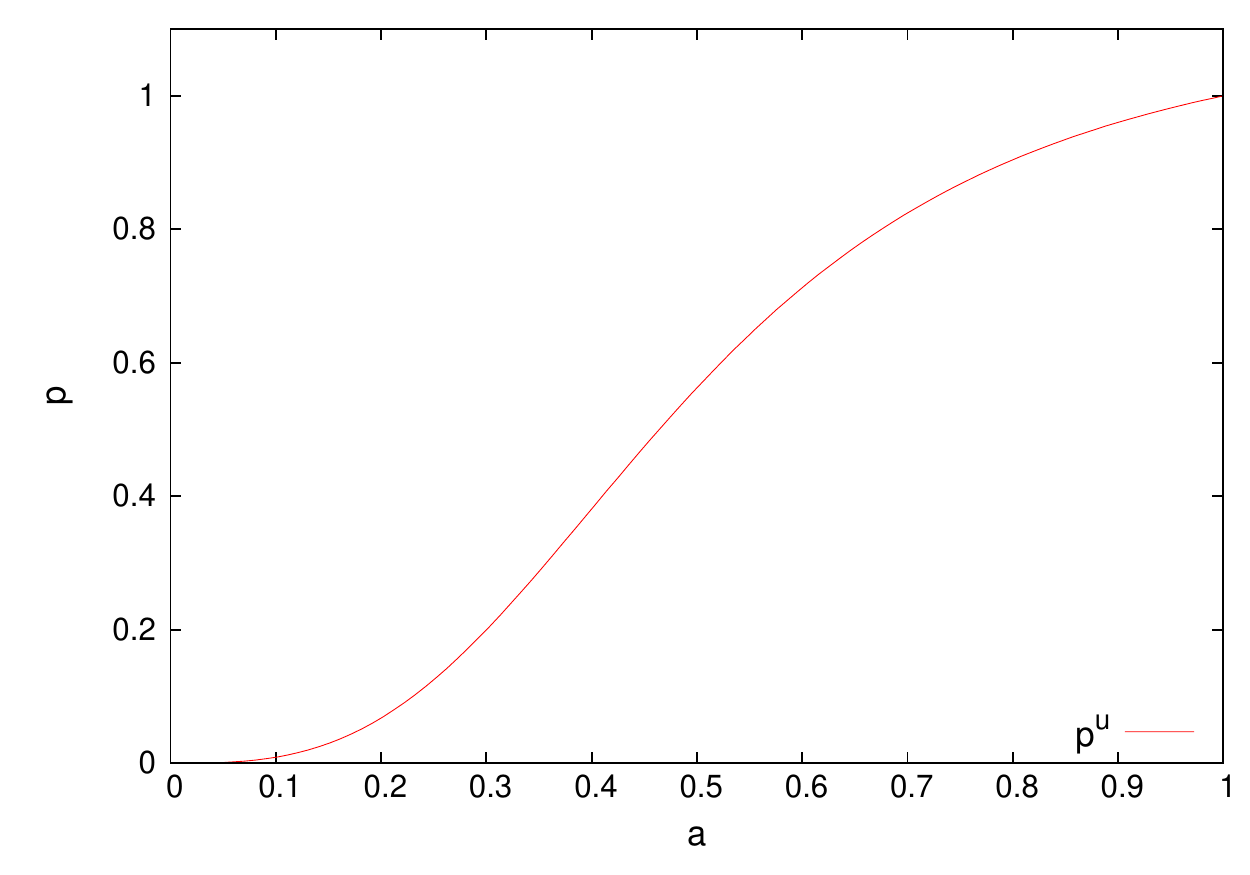}
\caption{The value of $p^U$ as a function of a}{In this figure, $a
= (\frac{y}{y-1})^{\frac{1}{3}}$. So when a goes from 0 to 1, y
goes from 0 to $\infty$. Note that as y goes to infinity, a goes to 1. We express the value as a function of a because it will be easier for us to combine different graphes into one graph later.}
\label{fig:theorem.jpg}
\end{figure}
Proof: Take $x > 0$, suppose there are n failure branches, whose
probabilities are $p_{f_1}, p_{f_2}, \cdot, p_{f_n}$, and the
corresponding interference terms are $-y_1, -y_2, \cdot -y_n$, then we
have

\begin{eqnarray}
& p_s x - \sum p_{f_i} y_i =0\\
& p_s + \sum p_{f_i} = 1
\end{eqnarray}

Rewrite it in the following form,
\begin{eqnarray}
& p_s x - p_{ft} y' =0\\
& p_s + p_{ft} = 1
\end{eqnarray}
where $p_{ft} = \sum p_{f_i}$ and $y' = \frac{\sum p_{f_i}
y_i}{p_{ft}}$. The solution of it is
\begin{equation}
p_s = \frac{y'}{x+y'}
\end{equation}
As the interference term of all the failure branches are larger than -y,
we have $y' < y$, then we can get $p_s < p^U(-y) =
\frac{y}{x+y}$. The discussion for the case when $x < 0$ is similar.
\QED

\begin{remark}
Recall the range of the I can be $-\infty < I \leq 1$, which means I can be unbounded below. Then in the $x > 0$ case, if the I of the failure branch goes to $-\infty$, or we can say y goes to $\infty$, we will have $p^U(-y)$ arbitrary close to 1. Therefore, theorem 4 alone is not enough for establishing a non-trivial upper bound. To derive a non-trivial upper bound, we need to find some additional constraints which are related to the interference term. In fact, this is what we will do in section 4.
\end{remark}

\section{Upper Bound for a general case}
\label{sec:upperbound2}

In this section, we will find an upper bound in a more general case.
Recall the problem of theorem 4  is that the interference can be
unbounded below. So we would like to find an additional constraint.
It turns out that the fact that the 3-tangle, a measure of
tripartite entanglement introduced in \cite{Coffman2000}, is an entanglement monotone (i.e., it cannot
increase on average under LOCCs) is precisely what we need
\cite{Dur2000}.

Our strategy is that, for any possible transformation protocol, we
would like to construct a new protocol that has the following two properties: 1. It has an
upper bound for the maximal successful probability of transformation
which is obviously smaller than one; 2. We can reconstruct the
original protocol from this new protocol, which means the successful
probability of this new protocol can be no less than the original
one. The way we construct such a protocol is given in 4.2 and the
bound of it will be given in 4.3, in which we deal with a special
example: the transformation from GHZ state to a special GHZ class
state $\ket{\phi} = \gamma(\ket{000} + \ket{aaa})$. In 4.4, we will
generalize this bound to more general cases, where we find
for any transformation from one GHZ-class state $\ket{\phi}$ to another GHZ-class state
$\ket{\psi}$ with different interference terms, we can find an nontrivial upper
bound for the successful probability.

\subsection{interference term and the maximal value of the 3-tangle of a
GHZ-class state}
Now consider such a question: Suppose we have an unknown GHZ class state
$\ket{\phi_{GHZ}} = \sqrt{K}(c_{\delta}\ket{0}\ket{0}\ket{0} +
s_{\delta}e^{i\varphi}\ket{\varphi_A}\ket{\varphi_B}\ket{\varphi_C})$  with a
given interference term f, what is the maximal value of the 3-tangle
$\tau_{ABC}$ \cite{Dur2000} ?

\begin{theorem}
\label{theorem: interference}
For a GHZ class
$\ket{\phi_{GHZ}}$, if its interference term is I, then the maximal value of
its 3-tangle is $\frac{(1 - a^2)^3}{(1 + a^3)^2}$, where a =
$(\frac{f}{1 - f})^{\frac{1}{3}}$.
\end{theorem}

The proof will be given in the appendix.

\subsection{"stop and reconstruct" procedure}

From \cite{Oreshkov2005}, we know every measurement can be seen as
constructed by many infinitesimal steps of weak measurement, that is, a
measurement which only slightly changes the original state. From this view, to get a better understanding of the transformation protocol, we would like to try to reduce the case where a failure branch gives an $I > I_0$ (some prescribed value) to the case where an undecided branch has $I = I_0$. That is to say, we are using a reduction idea. First we need to answer the
following question: Can we stop at some intermediate point and reconstruct
the original measurement? It turns out that the answer is yes. In fact, from \cite{Oreshkov2005}, the following theorem follows easily.

\begin{figure}[htp]
\centering
\includegraphics[width=0.5\textwidth, height=0.33\textwidth]{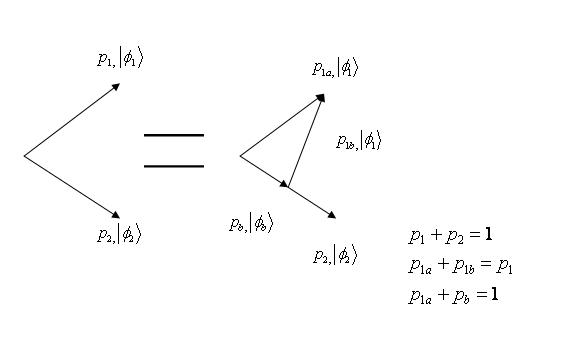}
\caption{"stop and reconstruct" for a two-outcome measurement}
\label{fig:stopandreconstruct2.jpg}
\end{figure}

\begin{theorem}
A two-outcome measurement $\{M_1, M_2\}$ can be reconstructed by stopping at
an immediate step $\{\sqrt{1-e^{-2x}} M'_1, \sqrt{1+e^{-2x}} M'(x)\}$ and a
reconstructing measurement ${\{M'(x,
+\infty), M'(x, -\infty)\}}$, where $M'_1 = \sqrt{M^+_1M_1}$ and
\begin{eqnarray}
M'(x, +\infty)=& \sqrt{\frac{1+\tanh(x)}{I+\tanh(x)(M'^2_2-M'^2_1)}}M'_2,\\
M'(x, -\infty)=& \sqrt{\frac{1-\tanh(x)}{I+\tanh(x)(M'^2_2-M'^2_1)}}M'_1
\end{eqnarray}
See Figure~\ref{fig:stopandreconstruct2.jpg} for a graphic discription.
\end{theorem}

Proof: Firstly, from polar decomposition we have $M_1=
U_1M'_1, M_2=U_2M'_2$, where $U_1$ and $U_2$ are unitary, $M'_2=
\sqrt{M^+_2M_2}$. Then
$\{M'_1, M'_2\}$ is also a measurement. As $M'_1$ and $M'_2$ are
positive, it can be reconstructed from infinitesimal steps
.\cite{Oreshkov2005} Secondly, instead of measure $\{M'_1, M'_2\}$, we
stop at $M'(x)= \sqrt{\frac{I + \tanh(x)(M'^2_2 - M'^2_1)}{2}}$ before
we reach $M'_2$, that is to say, we perform measurement
$\{\sqrt{1-e^{-2x}} M'_1, \sqrt{1+e^{-2x}} M'(x)\}$. The effect is
we still got $M'_1\rho M'^+_1/p_1$ but the probability become
$\sqrt{1-e^{-2x}}p_1$, but instead of get $M'_2\rho M'^+_2/p_2$, we
get $M'(x)\rho M'^+(x)/p(x)$ where $p(x)= Tr(M'(x)\rho M'^+(x))$.
Thirdly, we do nothing to the $M'_1$ branch, but do a POVM ${\{M'(x,
+\infty), M'(x, -\infty)\}}$.

On the M'(x) branch, it is easy to prove that,
\begin{equation}
M'(x, \infty)M'(x)= e^{-x}M'_1, M'(x, -\infty)M'(x)= M'_2,
\end{equation}
So in total, we perform a POVM $\{\sqrt{1-e^{-2x}}M'_1, e^{-x}M'_1, M'_2\}$,
that is just the same as $\{M'_1, M'_2\}$. Finally, if we get the result of
measurement $M'_1 (M'_2)$, perform a unitary transformation $U_1 (U_2)$, we
can reconstruct $\{M_1, M_2\}$ with a stop in the middle. QED.

However, a protocol may contain many measurements
and measurements with more than two outcomes, can we still use this method
to stop in the middle and reconstruct everything?

\begin{figure}[htp]
\centering
\includegraphics[width=0.5\textwidth, height=0.33\textwidth]{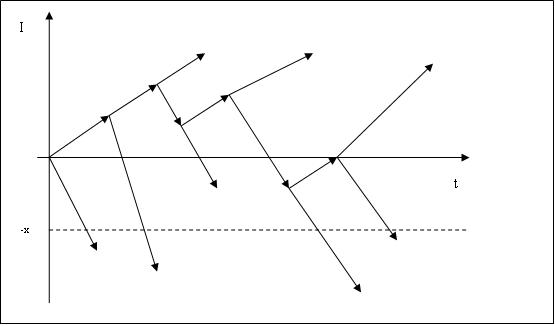}
\caption{The original protocol written in the many two-outcome measurements form}
\label{fig:originalprotocol.jpg}
\end{figure}

\begin{figure}[htp]
\centering
\includegraphics[width=0.5\textwidth, height=0.33\textwidth]{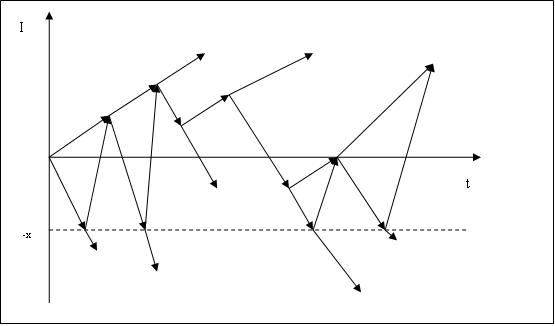}
\caption{"stop and reconstruct" for general protocol, I stands for the interference term}
\label{fig:stopandreconstructn.jpg}
\end{figure}

The answer is yes. To show this, first we need to rewrite every measurement
in the protocol into a sequence of two-outcome measurements
\cite{Andersson2007}, see Figure~\ref{fig:originalprotocol.jpg}. Then the protocol consists of only two-outcome
measurements. So the "stop and reconstruct" can work for each of them. The
only thing is that, now, each two-outcome measurement may be related to many
other two-outcome measurements, so during the "stop and reconstruct"
process, many measurements might be affected. How can we be sure we can
reconstruct everything? For this problem, notice that these two-outcome
measurements are all in order. Then when we do the "stop and reconstruct",
the principle is that we should always stop at the earlier two-outcome
measurement first. Moreover, we need to reconstruct the earlier ones first. See Figure~\ref{fig:stopandreconstructn.jpg}.

\subsection{Example: $\ket{GHZ} \rightarrow \ket{\phi} = \gamma(\ket{000} + \ket{aaa})$}

Now, we want to find an upper bound for the success probability of the
transformation.

\begin{theorem}
Suppose we have a SLOCC transformation protocol from $\ket{GHZ}$ to
$\ket{\phi} = \gamma(\ket{000} + \ket{aaa})$, where $\ket{a}=
c\ket{0}+ \sqrt{1-c^2}\ket{1}$ and $c \in (0,1]$. Suppose the
successful probability is $p_m$. Then we can always find a protocol
consisting of only successful
and failure branches which has a successful probability no less than $p_m$.
\end{theorem}

Proof: If the protocol is in that form, we do nothing. If the
protocol has some branches which are neither successful nor failure. Then we do nothing to the successful or failure branches. However, for the undecided branches, from the definition of it we know we can always find a POVM that can transform it into the desired state with nonzero probability $\delta_p$. Then the total successful probability is $p_m + \delta_p$, which is higher than $p_m$. In all, we can always find a protocol consisting of only successful and failure
branches which have a successful
probability no less than $p_m$. QED.

\begin{figure}[htp]
\centering
\includegraphics[width=0.5\textwidth, height=0.33\textwidth]{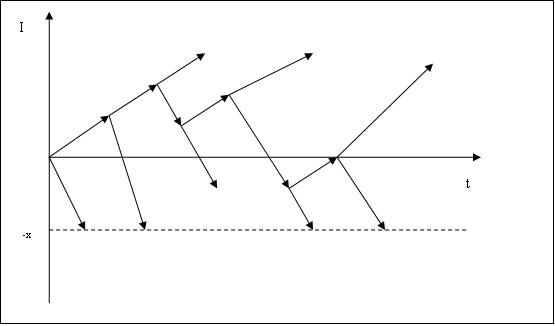}
\caption{The new protocol, which can reconstruct the original one}
\label{fig:newprotocol.jpg}
\end{figure}

Now modify the protocol we get in the first step in the following way:

Suppose we can find at least one failure branch that
have interference term smaller than -y, where ${y \geq 0}$. then we can
find a x, where ${0\leq x \leq y}$. As our initial interference term is zero,
now we can use the weak measurement idea to let all the branches stop if its
interference term reaches -x and do nothing to the branches which never reach
-x. And we can get a new protocol in Figure~\ref{fig:newprotocol.jpg}

\begin{remark}
Note that to make this new protocol work, we have applied the intermediate value theorem. That is to say, we implicitly assume that
the interference terms, I, of the two intermediate states specified in Theorem 5,
are continuous functions of x. This assumption works because, from \cite{Oreshkov2005},
we know $M(x, \delta x)$ changes the state given by $M(x)\ket{GHZ}$ very little, or we can say it is a weak measurement. While from the expression of interference term Equation~\eqref{eq:interferencetermdefi} in Appendix, we know interference term is a continuous function of the parameters of the state. Then, as the state changes very little under the weak measurement, the interference term also changes continuously.
\end{remark}

Then we get a new protocol. It has two properties:

1) There are three kinds of branches: failure branches with interference term
larger than or equal to -x, successful branches and the branches neither
successful nor failure with interference term -x.

2) From the "stop and reconstruct" part, we know we can reconstruct
the original protocol by performing LOCCs (may be a sequence of
measurements) just on these branches which have interference term -x and
do nothing on other branches. That is to say, just do LOCCs on the
-x branches, we can get a total successful probability no less than
the original one. So, if we have an upper bound of successful
probability for the new protocol, that should also be an upper bound
for the original one.

Then we can find the upper bound for this new protocol. Now, the
protocol consists of three kinds of branches: successful branches,
failure branches with interference term larger than or equal to -x,
and undecided branches with interference term -x. The total
successful probability of this protocol consists of two probability:
the already existing successful branches' total probability $p_{se}$
and the probability we can transform from the -x branches to the
states we want.

\begin{theorem}
\label{theorem:newprotocol} As in Theorem 7, we consider a SLOCC
transformation from $\ket{GHZ}$ to $\ket{\phi} = \gamma(\ket{000} +
\ket{aaa})$. For all the possible new protocols shown in
Figure~\ref{fig:newprotocol.jpg}, there is an upper bound for the
success probability
\begin{equation}
\bar{p_s}(-x)= p_{as}(-x) +p_u(-x)*p_m(s|-x)
\end{equation}
where $p_{as}(-x)$ is the already successful branches in this
condition, while $p_u(-x)$ is the probability of the undecided
branches with interference term -x. And $p_m(s|-x)$ is the maximal
probability to transform a GHZ-class state with interference term -x
into the destination state $\phi_s$. And we will get

$p_{as}= \frac{\frac{a^3}{1-a^3}}{\frac{c^3}{1+c^3} +
\frac{a^3}{1-a^3}}$, $p_u(-x) = 1 - p_{as}$,

$p_m(s|-x) = min(\frac{max(\tau_{ABC}(\phi | I(\phi) =
-x))}{\tau_{ABC}(\phi_s)},1)$

where a is the solution of the equation $x = \frac{a^3}{1-a^3}$,
$\tau_{ABC}$ stands for the 3-tangle.
\end{theorem}

We firstly consider the case that there exists no failure branches
with interference term larger than -x, later we will show the other
case can only give an upper bound smaller than in this case.

\begin{lemma}

\label{lem: upperbound} If the new protocol shown in Figure~\ref{fig:newprotocol.jpg}
consists of only
successful branches and branches with interference term -x (no
failure branches with interference term larger than -x), it has an
upper bound
\begin{equation}
\bar{p_s}(-x)= p_{as}(-x) +p_u(-x)*p_m(s|-x)
\end{equation}
where the parameters are defined in
Theorem~\ref{theorem:newprotocol}.
\end{lemma}

Proof: For the already existing successful branches, the total probability
is determined by -x and the conservation of interference term. As $x = \frac{a^3}{1-a^3}$, we have

\begin{eqnarray}
& p_{as}(-x)I(\ket{\phi_s}) - p_u(-x)x = 0\\
& p_{as}(-x) + p_u(-x) =1
\end{eqnarray}

Solving Eqs. (66) and (67), we can find $p_{as}(-x)=
\frac{\frac{a^3}{1-a^3}}{\frac{c^3}{1+c^3} + \frac{a^3}{1-a^3}}$.

For the maximum value of $p_s(-x)$, using the 3-tangle idea, we know
it is bounded by $p_u(-x)*p_m(s|-x)$ where
\begin{equation}
p_m(s|-x) =
min(\frac{max(\tau_{ABC}(\phi | I(\phi) = -x))}{\tau_{ABC}(\phi_s)},1)
\end{equation}

Then we find an upper bound for the successful probability of this
new protocol when there is no failure branch having interference term larger than
-x:
\begin{equation}
\bar{p_s}(-x)= p_{as}(-x) +p_u(-x)*p_m(s|-x)
\end{equation}
\QED

\begin{remark}
To show it is really an
upper bound for the successful probability for the new protocol,
we need to show if there is any other failure branch with interference term larger
than -x,
we can only get a successful probability smaller than this.
\end{remark}

Proof of Theorem~\ref{theorem:newprotocol}: To prove this theorem,
we just need to prove the following: If the new protocol contains a
failure branch which has an interference term larger than -x, it has
an upper bound for the success probability smaller than what we get
in Lemma~\ref{lem: upperbound}.

Consider the conservation of interference term, now we have:

\begin{eqnarray}
& p'_{as}(-x)I(\ket{\phi_s}) + \sum p_{fi}I(\ket{\phi_{f_i}}) - p'_u(-x)x = 0\\
& p'_{as}(-x) + \sum p_{fi} + p'_u(-x) =1
\end{eqnarray}

which can be rewritten as from Corollary 2

\begin{eqnarray}
& p'_{as}(-x)I(\ket{\phi_s}) - p"(-x')x' = 0\\
& p'_{as}(-x) + p"(-x') =1
\end{eqnarray}

where

\begin{eqnarray}
& x' = \frac{\sum p_{fi}I(\ket{\phi_{f_i}}) - p'_u(-x)x}{\sum p_{fi} +
p'_u(-x)}\\
& p"(-x') = \sum p_{fi} + p'_u(-x)
\end{eqnarray}

As $I(\ket{\phi_{f_i}})>-x$, we know $-x'>-x$, so $p'_{as}(-x)< p_{as}(-x)$. Let the difference between $p'_{as}(-x)$ and $p_{as}(-x)$ be $\delta_s$, then we have
$\delta_s = p_{as}(-x) - p'_{as}(-x)$, so $\delta_s > 0$ and $p"(-x')=
p_u(-x) + \delta_s$. As
$\sum p_{fi} > 0$, we have $p'(-x) < p"(-x') = p_u(-x) + \delta_s$. So
the total successful probability in this case is

\begin{eqnarray}
& \bar{p'_s} =  p'_{as}(-x) + p'(-x)*p(s|-x)\notag\\
      \leq &p'_{as}(-x) + p'(-x) * p_m(s|-x)\notag\\
      < & p_{as}(-x) - \delta_s + (p_u(-x) + \delta_s) * p_m(s|-x)\notag\\
      < & p_{as}(-x) + p_u(-x) * p_m(s|-x)\notag\\
        & + \delta_s(p_m(s|-x) - 1)\notag\\
      \leq & p_{as}(-x) + p_u(-x) * p_m(s|-x)\notag\\
      = & \bar{p_s}(-x)
\end{eqnarray}

Here, in the second last step, we have used the fact that $p_m(s|-x)$ can not be larger than one.
So we know it is really an upper bound for the successful
probability for the new protocol, which should also be an upper
bound for the successful probability for the original protocol, and
an upper bound for the transformation protocols which contains at
least on failure branch which has interference term smaller than -x
( or we can say it passes -x). So we have $\bar{p'_s} <
\bar{p_s}(-x)$, which means $\bar{p_s}(-x)$ is an upper bound for
the new protocol. \QED

\begin{corollary}
As in Theorem 7, we consider a SLOCC transformation from $\ket{GHZ}$
to $\ket{\phi} = \gamma(\ket{000} + \ket{aaa})$. If a protocol
contains at least one failure branch whose interference term is
smaller than -y, its successful probability should be bounded by all
the $\bar{p_{s}}(-x)$, where $0 \leq x \leq y$.
\end{corollary}

Proof: If we see every branch from the weak measurement idea. We
will find the interference term should change continuously, so we
can stop at any point between 0 and -y. For each point we choose, we
can get an upper bound. And all the upper bounds should be the upper
bounds of the original branch. \QED

\begin{corollary}
For a LOCC transformation protocol from $\ket{GHZ}$ to $\ket{\phi} =
\gamma(\ket{000} + \ket{aaa})$, if the minimum interference term of
all the failure branches is -z, then its successful probability
should be bounded by

\begin{equation}
p_{bound}(-z) = \min(p^U(-z), \bar{p}_{\tau_{ABC}}(-z))
\end{equation}

where $\quad \bar{p}_{\tau_{ABC}}(-z)) = \min_{0 \leq x \leq
z}(\bar{p_{s}}(-x))$.
\end{corollary}

\begin{figure}[htp]
\centering
\includegraphics[width=0.5\textwidth, height=0.33\textwidth]{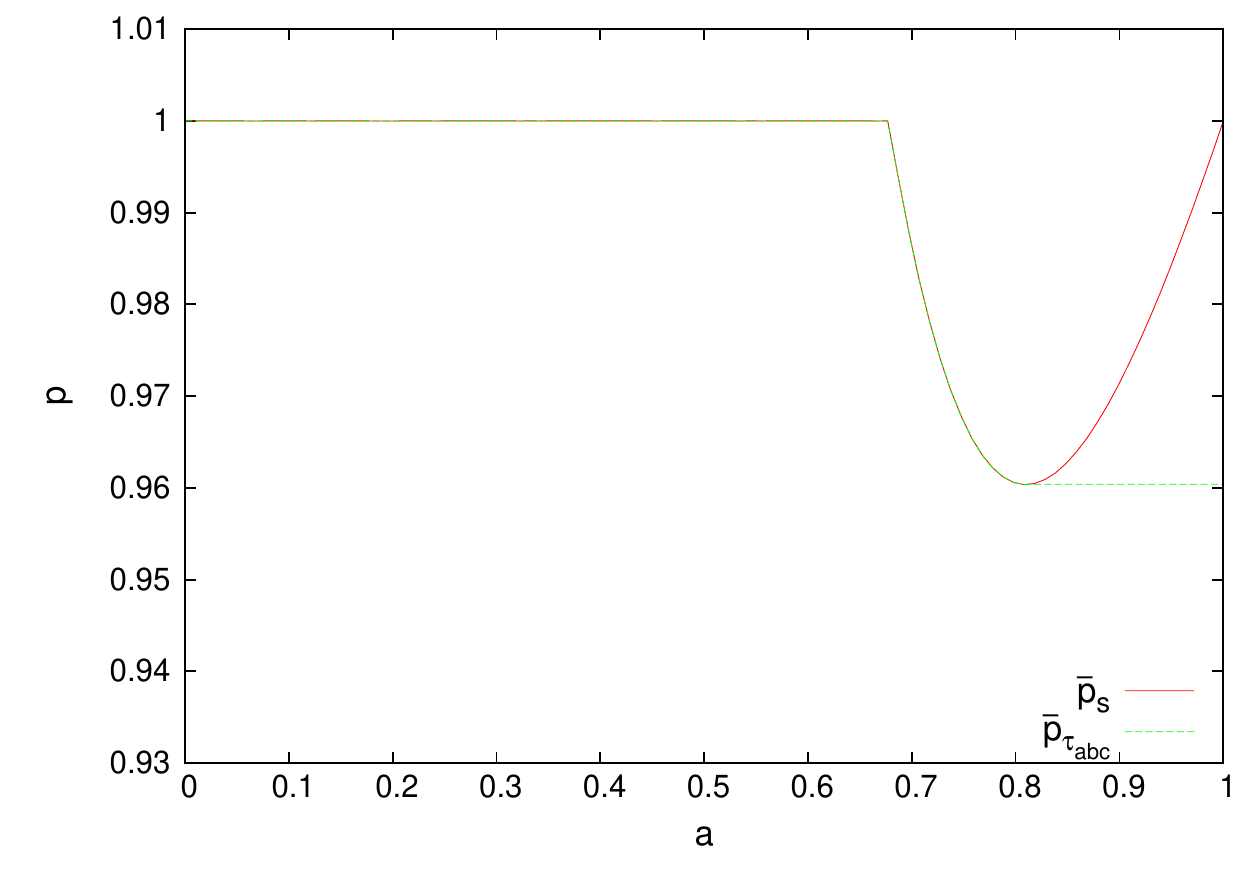}
\caption{the relation between $\bar{p_s}$ and $\bar{p}_{\tau_{ABC}}$}
\label{fig:3tanglebound.jpg}
\end{figure}

Proof: If the minimum interference term is -z, then from
Theorem~\ref{theorem:failurebound}, we know there is an upper bound
$p^U(-z) = \frac{z}{I(\ket{\phi_s})+z}$, which is in fact
$p_{as}(-z)$. As it is bounded by all the $\bar{p_{s}}(-x)$, where
$0 \leq x \leq z$, we can find another upper bound
$\bar{p}_{\tau_{ABC}}(-z)) = \min_{0 \leq x \leq
z}(\bar{p_{s}}(-x))$. See Figure~\ref{fig:3tanglebound.jpg} for the
relation between $\bar{p_s}$ and $\bar{p}_{\tau_{ABC}}$. Then the
minimum of these two bounds is also an upper bound, which we call
$p_{bound}(-z)$.

\begin{figure}[htp]
\centering
\includegraphics[width=0.5\textwidth, height=0.33\textwidth]{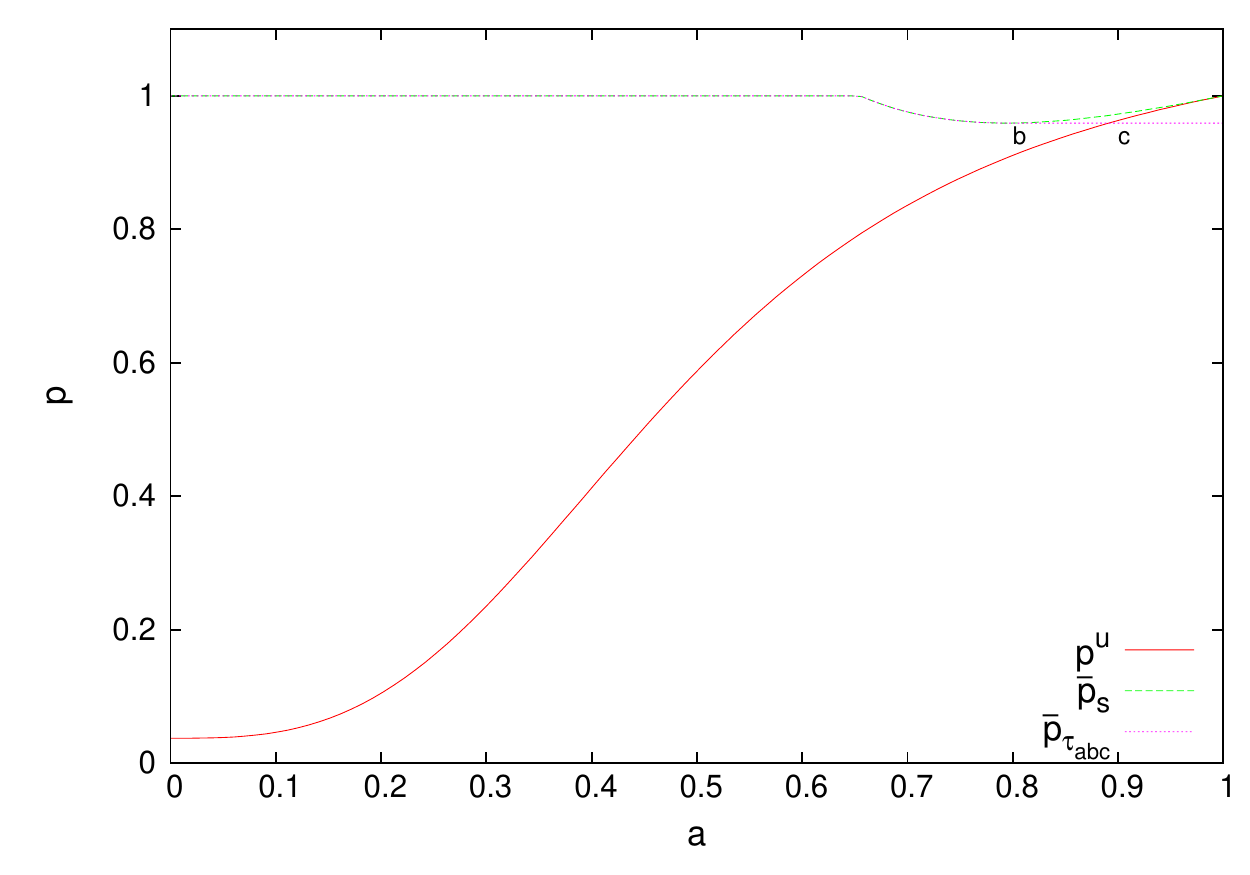}
\caption{The upper bound for the transformation}{In this figure, $a
= (- \frac{x}{x-1})^{\frac{1}{3}}$. So when a goes from 0 to 1, x
goes from 0 to $\infty$. The dashed line is the plot of $\bar{p_s}$ as
a function of -x, the dot line is the plot of $\bar{p}_{\tau_{ABC}}$, the solid line is
the plot of $p^U$. Notice that point b corresponds to the minimum
value of $\bar{p_s}$, before b, $\bar{p_s}$ decreases
monotonically. So before b, $\bar{p}_{\tau_{ABC}}$ is the same as
$\bar{p_s}$, while after b, $\bar{p}_{\tau_{ABC}}$ remains to be
the value of $\bar{p_s}$ at b. Another thing is that before c,
$p^U$ is smaller than $\bar{p}_{\tau_{ABC}}$, while after c,
$\bar{p}_{\tau_{ABC}}$ is smaller than $p^U$. So the final plot we get
for the upper bound is the solid line before c and the dot line after c, which we call upper bound line. The meaning of this upper bound
line is that, for a given transformation protocol, if the smallest
interference term of all the failure branches is -y, let $k = (-
\frac{y}{y-1})^{\frac{1}{3}}$, the success probability can not be
larger than the corresponding point in the upper bound line. Consider all of
the possible protocols (a goes from 0 to 1), the upper bound of the
transformation probability is the largest value of the points on
the upper bound line, which is just the minimum value of $\bar{p_s}$}
\label{fig:bounds.jpg}
\end{figure}

\begin{theorem}
An upper bound of LOCC transformation from GHZ state to a specific
GHZ class state $\ket{\phi} = \gamma(\ket{000} + \ket{aaa})$ is the
maximum value of  $p_{bound}(-z)$£¬ where $z \in [0, +\infty)$. And
it is in fact the minimum value of $\bar{p_s}(-z)$, where $z \in [0,
+\infty)$.
\end{theorem}

Proof: The basic picture of our proof can be represented in
Figure~\ref{fig:bounds.jpg}.

Now we consider all the possible transformation protocols. Then the value of the minimum interference term -z may vary from 0 to $\infty$. (We can always find a protocol giving a very small value of -z, while its successful probability is still bounded.) Easy to see an upper bound is the maximum value of $p_{bound}(-z)$ for all the possible values of z, where $z \in [0, \infty)$.  In fact, we can find the upper bound we get
for this transformation is the minimum value of $\bar{p_{s}}(-z)$,
where $z \in [0, +\infty)$. \QED

Put $\tau_{ABC}(\phi)= \frac{(1-c^2)^3}{(1+c^3)^2}$, in to the
equation, we can get the upper bound. The analytic value is hard to
get, if we put c=0.5. The minimum value of $\bar{p_s}(-x)=
p_{as}(-x) + p_u(-x)*p_m(s|-x)$ is 0.9604 at x=1.13062, which is
less than 1.

\subsection{general case}

In the above, we have considered an upper bound for the special case of $\ket{GHZ} \rightarrow
\ket{\phi} = \gamma (\ket{000} + \ket{aaa})$ to find the upper bound
for it. Now, we will consider two more general cases. First, we will consider the transformation $\ket{GHZ} \rightarrow \ket{\phi_{GHZ}} =
\sqrt{K}(c_{\delta}\ket{0}\ket{0}\ket{0} +
s_{\delta}e^{i\varphi}\ket{\varphi_A}\ket{\varphi_B}\ket{\varphi_C})$,
which is the general GHZ class state; Second, we will consider a general GHZ class state
to another general GHZ class state.

1. $\ket{GHZ} \rightarrow \ket{\phi_{GHZ}} =
\sqrt{K}(c_{\delta}\ket{0}\ket{0}\ket{0} +
s_{\delta}e^{i\varphi}\ket{\varphi_A}\ket{\varphi_B}\ket{\varphi_C})$.
In this case, we just need to change the expression for the interference term and 3-tangle of the destination state into

\begin{eqnarray}
& Interference \; term: I(\phi_s) =
\frac{2c_{\alpha}c_{\beta}c_{\gamma}s_{\delta}c_{\delta}c_{\varphi}}{(1+2c_{
\alpha}c_{\beta}c_{\gamma}s_{\delta}c_{\delta})}\\
& 3 - tangle: \tau_{ABC}(\phi_s) =
\frac{4s^2_{\alpha}s^2_{\beta}s^2_{\gamma}s^2_{\delta}c^2_{\delta}}{(1+2c_{
\alpha}c_{\beta}c_{\gamma}s_{\delta}c_{\delta})^2}
\end{eqnarray}

Then we can use the similar process, except changing the corresponding value of Interference term and 3-tangle, see the following for details.

Firstly, using the "stop and reconstruct" method to get the new protocol with only successful branches and undecided branches with interference term x. We have

\begin{eqnarray}
p_{as}(x)I(\ket{\phi_s}) + p_u(x)x &= 0\\
p_{as}(x) + p_u(x) &=1
\end{eqnarray}

We have
\begin{eqnarray}
p_{as}(x) &= \frac{x}{x-I(\ket{\phi_s})} = 1 - p_u(x)\\
p_m(s|x) &= \min(\frac{max(\tau_{ABC}(\phi | I(\phi) = x))}{\tau_{ABC}(\phi_s)},1)
\end{eqnarray}

Then, the supremum success probability of this new protocol should be bounded by
\begin{equation}
\bar{p_{s}}(x)= p_{as}(x) +p_u(x)*p_m(s|x)
\end{equation}

Consider all possible protocols, we find the minimum of $\bar{p_{s}}(x)$ where $x\in [0, \frac{1}{2}]$ if $I(\ket{\phi_s}) < 0$ and $x\in (-\infty, 0]$ if $I(\ket{\phi_s}) > 0$ is an upper bound of the success probability of this transformation.

\begin{remark}
Suppose we want to transform a GHZ state to a
GHZ-class state $\ket{\phi_{GHZ}} =
\sqrt{K}(c_{\delta}\ket{0}\ket{0}\ket{0} +
s_{\delta}e^{i\varphi}\ket{\varphi_A}\ket{\varphi_B}\ket{\varphi_C})$. If $I(\ket{\phi_{GHZ}}) \neq 0$, we can always find a nontrivial upper bound. However, for the case where $I(\ket{\phi_{GHZ}}) = 0$, we will get a trivial upper bound 1. This condition consists of 2 possibilities: 1. $\braket{000|\varphi_A\varphi_B\varphi_C} =
0$; 2. $\varphi = \frac{\pi}{2}$ or $\frac{3\pi}{2}$. In fact, in the paper
\cite{Turgut2009}, they have provided a protocol for such a transformation with success probability 1.
\end{remark}

2. A general GHZ class state to another general GHZ class state. In
this case, the interference term is still conserved, but the initial
value should be the interference term of the initial state.

\begin{eqnarray}
\label{eq:palready}
p_{as}(x)I(\ket{\phi_s}) + p_u(x)x &= I_{initial}\\
p_{as}(x) + p_u(x) &=1
\end{eqnarray}

We have
\begin{eqnarray}
p_{as}(x) &= \frac{I_{initial}-x}{I(\ket{\phi_s}) - x} = 1 - p_u(x)\\
p_m(s|x) &= \min(\frac{max(\tau_{ABC}(\phi | I(\phi) = x))}{\tau_{ABC}(\phi_s)},1)
\end{eqnarray}

Then, the supremum success probability of this new protocol should be bounded by
\begin{equation}
\bar{p_{s}}(x)= \bar{p_{s}}(x) +p_u(x)*p_m(s|x)
\end{equation}

Consider all possible protocols, we find the minimum of $\bar{p_{s}}(x)$ where $x\in [I_{initial}, \frac{1}{2}]$ if $I(\ket{\phi_s}) < I_{initial}$ and $x\in (-\infty, I_{initial}]$ if $I(\ket{\phi_s}) > I_{initial}$ is an upper bound of the success probability of this transformation.

\begin{example}
An upper bound for the transformation from $\ket{\phi} =
\gamma(\ket{000} + \ket{abc})$  where $\braket{0|a} = 0.1$,
$\braket{0|b} = 0.2$, $\braket{0|c} = 0.2$, to $\ket{\psi} =
\gamma'(\ket{000} + \ket{a'b'c'})$ where $\braket{0|a'} = 0.4$,
$\braket{0|b'} = 0.5$, $\braket{0|c'} = 0.6$ is 0.9593. See
Figure~\ref{fig:upperboundexp2.jpg}.
\begin{figure}[htp]
\centering
\includegraphics[width=0.5\textwidth, height=0.33\textwidth]{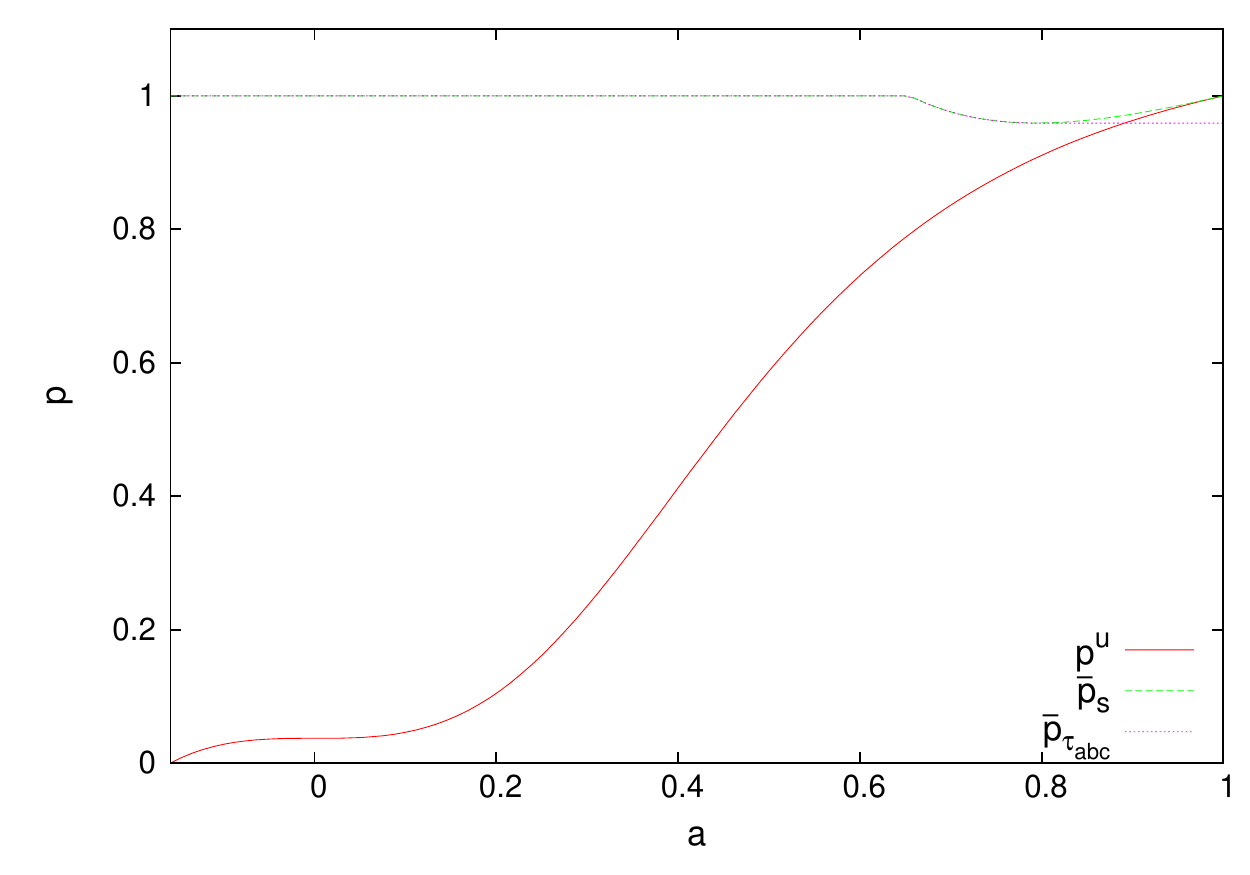}
\caption{upper bound of transformation probability from $\ket{\phi}$ to $\ket{\psi}$}
\label{fig:upperboundexp2.jpg}
\end{figure}
\end{example}

\begin{lemma}
For a transformation from a tripartite state $\ket{\phi_i}$ to another tripartite state
$\ket{\phi_s}$,  if the interference term of $\ket{\phi_i}$ is not equal
with $\ket{\phi_s}$, we can get an upper bound for the supremum of
the successful probability which is less than 1.
\end{lemma}

Proof: If the interference term is not equal, put into the Equation~\ref{eq:palready}
, we know the maximal successful probability
$p_{as}(x)$ can not reach one. Otherwise the conservation of
interference term will be violated. In fact, as the magnitude of the
interference term of branches where we stop become larger and
larger, $p_{as}(x)$ gets closer and closer to one. However, at
the same time the maximal 3-tangle of these branches will go to
zero. As the destination state is in GHZ class, its 3-tangle is not
zero. So we can always find one $|x|$, when the magnitude square of
the interference term reaches it, $max(\tau_{ABC}(\phi |
|I(\phi)| = |x|)) < \tau_{ABC}(\phi_s)$, then it will give an
upper bound for this transformation which is smaller than 1. \QED

\begin{remark}
One may naturally ask a question: If the interference terms of two states are the same, can we give an upper bound for the transformation probability? In this case, the above lemma can not give a nontrivial upper bound. However, we can still use other entanglement monotones, such as 3-tangle, to give an upper bound for the transformation from one to another.
\end{remark}

\begin{example}
Consider the transformation from $\ket{\phi} = \gamma(\ket{000} +
\ket{abc})$ where $\braket{0|a} = 0.2$, $\braket{0|b} = 0.4$,
$\braket{0|c} = 0.8$, to $\ket{\psi} = \gamma'(\ket{000} +
\ket{a'b'c'})$ where $\braket{0|a'} = 0.4$, $\braket{0|b'} = 0.4$,
$\braket{0|c'} = 0.4$. One can check that $I(\ket{\phi}) =
I(\ket{\psi}) = 0.0602$. So naively we can only get a trivial upper
bound for the transformation between them. However, notice that
$\tau_{ABC}(\ket{\phi}) = 0.2564$ and $\tau_{ABC}(\psi) = 0.5235$,
we can get an upper bound for the transformation from $\ket{\phi}$
to $\ket{\psi}$ which is

\begin{equation}
\frac{\tau_{ABC}(\ket{\phi})}{\tau_{ABC}(\ket{\psi})} =
\frac{0.2546}{0.5235} = 0.4863 < 1
\end{equation}

\end{example}

\section{Lower Bound for the Transformation}
\label{sec:lowerbound}

After the discussion about the upper bound, we have a question: Is
this upper bound tight or not? Or can we find a transformation
protocol that can reach such a transformation probability? In
\cite{Chitambar2008}, they have provided a straight forward
protocol: That is, if we want to transform

\begin{eqnarray}
&\GHZ = \frac{1}{\sqrt{2}} (\ket{000} + \ket{111}) \notag\\
\stackrel{\textrm{LOCC}} {\longrightarrow} \ket\Psi &=
\sqrt{K}(\ket{000} + \ket{\varphi_A}\ket{\varphi_B}\ket{\varphi_C}),
\end{eqnarray}
We can let Alice perform the measurement $\{\frac{A}{\|A\|},
\sqrt{\Id - \frac{1}{\|A\|^2}A^+A}\}$, where
\begin{equation}
A\ket{0} = \ket{0}, A\ket{1} = \ket{\varphi_A}
\end{equation}
and similarly for Bob and Charlie. Then the final successful
probability is
$\frac{(1+c_{\alpha}c_{\beta}c_{\gamma})^3}{(1+c_{\alpha})(1+c_{\beta})(1+c_{\gamma})}$.

In the following, we will provide a transformation protocol that can
transform GHZ-state generalized to n parties and m dimensions to
other states with the same dimension and Schmidt rank with a
probability higher than the straight forward protocol . However,
there is still a gap between the upper bound and the lower bound we
get. A surprising result derived from that is all tripartite pure
3-qubit states can be transformed from GHZ-state generalized to 3
parties and 3 dimensions by LOCC with probability 1, which was not
known before.

We have found a lower bound for the maximum value of the transformation probability from GHZ state to $\ket\Psi$= $\gamma(\alpha \ket{a_1b_1c_1} +\beta \ket{a_2b_2c_2})$ by an explicit method, which we call four-step method.

\begin{figure}[htp]
\centering
\includegraphics[width=0.5\textwidth, height=0.33\textwidth]{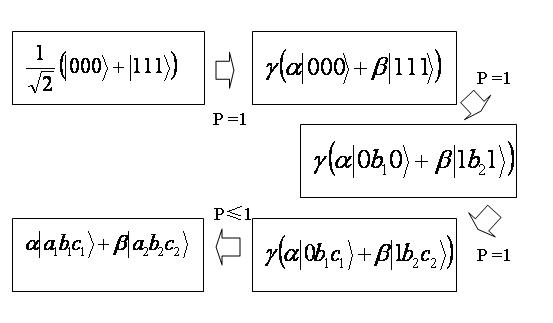}
\caption{Four-step method}
\label{fig:fourstepmethod.jpg}
\end{figure}

See Figure~\ref{fig:fourstepmethod.jpg} for the process, in the first step, we transform GHZ state into $\ket{GHZ'}=\alpha\ket{000}+\beta\ket{111}$ which has the same coefficient of terms (but different states) as $\ket\Psi$. Secondly, we transform $\ket{GHZ'}$ into $\ket{\phi_{b1}}= \alpha\ket{0b_10} + \beta\ket{1b_21}$. Then we transform $\ket{\phi_{b1}}$ into $\ket{\phi_{c1}} = \alpha\ket{0b_1c_1} + \beta\ket{1b_2c_2}$. We will show the first three steps can be done with probability 1. Then if $\braket{a_1|a_2}$ is zero, $\ket{\phi_{c1}}$ is unitary equivalent with $\ket{\Psi}$ so we can get $\ket{\Psi}$ with probability 1. In other cases, we can still get $\ket{\Psi}$ with a higher probability than the previous result in \cite{Chitambar2008} by performing an appropriate measurement.

In fact, the first step is a generalization of Nielsen Majorization
result \cite{Nielsen1999} and Lo-Popescu's \cite{Lo2001} result for
the maximum probability of distilling a maximally entangled state.
It has been noted previously in \cite{Xin2007}.

\begin{definition}
  A GHZ-like (aka Schmidt decomposable) state is a tripartite state that
can be written in the form
    \begin{equation}
      \ket\Psi = \sum_i \lambda_i \ket{i}\ket{i}\ket{i}
    \end{equation}
\end{definition}

Theorem 1 of Lo-Popescu also holds for the GHZ-like state, because it gives
a bound for the case where the Bob-Charlie alliance is allowed to perform
any (non-local) operations, and when the allowed operations are restricted
to the subclass of local operations, the upper bound still has to hold.

Theorem 2(a) of Lo-Popescu can in the same way be
applied to GHZ-like (Schmidt decomposable) states, because all unitary
transformations on Bob's side involve only a relabeling of the basis states
($\ket{i} \leftrightarrow \ket{j}$), and therefore extending it to GHZ-like
states just changes this step to ($\ket{i}\ket{i} \leftrightarrow
\ket{j}\ket{j}$), which can also be done by local unitaries only.

Theorem 2(b) generalizes to GHZ-like states as well, because here
Alice performs all the operations and Bob either has to perform no operation
on his state at all (result "success"), or he has to discard it completely
(result "failure"), which can also be done if Bob's state is distributed
among Bob and Charlie.

In \cite{Xin2007}, it was shown that Nielsen's majorization idea
generalized to more parties can be applied to GHZ-like state, which
means the transformation
\begin{gather}
\begin{split}
\ket{GHZmn}= \frac{1}{\sqrt{m}}
\sum^m_{i=1}\ket{i}_1\ket{i}_2...\ket{i}_n\\
\rightarrow \ket\Psi = \sum^m_{i=1} \alpha_i
\ket{i}_1\ket{i}_2...\ket{i}_n
\end{split}
\end{gather}
can success with probability 1.

For the second and third step, we have the following lemma.

\begin{lemma}
    \cite{Turgut2009} The GHZ state can be transformed to $\ket\Psi$= $\gamma(\alpha \ket{a_1b_1c_1} +\beta \ket{a_2b_2c_2})$, where $\braket{\Psi|\Psi} =1 $ with probability 1, if $\ket{a_1b_1c_1}$ and $\ket{a_2b_2c_2}$ are orthogonal to each other.
\end{lemma}

Proof:
Suppose $\ket{a_1}$ and $\ket{a_2}$ are orthogonal to each other. If we choose the basis in which $\ket{a_1b_1c_1}= \ket{000}$ then we write $\ket\phi= \gamma(\alpha\ket{000} + \beta\ket{1}_A(d_1\ket{0}+d_2\ket{1})_B(e_1\ket{0}+e_2\ket{1})_C)$, where $|d_1|^2+|d_2|^2=1$ and $|e_1|^2+|e_2|^2=1$, in this case we can see $\gamma=1$

Then we can do the transformation in the following way.

Firstly, use the result of the first step to transform $\ket{GHZ}$ into $\ket{GHZ'}=\alpha\ket{000}+\beta\ket{111}$ with probability 1.

Secondly, Bob performs a POVM
\begin{equation}
    M_1 = \frac{1}{\sqrt{2}}
    \lk
    \begin{array}{cc}
        1 & d_1\\
        0 & d_2
    \end{array}
    \rk,
    \qquad
    M_2 = \frac{1}{\sqrt{2}}
    \lk
    \begin{array}{cc}
        1 & -d_1\\
        0 & -d_2
    \end{array}
    \rk,
\end{equation}
Then with probability $\frac{1}{2}$ we get $\ket{\phi_{b1}}= \alpha\ket{000} + \beta\ket{1}_A(d_1\ket{0}+d_2\ket{1})_B\ket{1}_C$, and with probability $\frac{1}{2}$ we get $\ket{\phi_{b2}}= \alpha\ket{000} - \beta\ket{1}_A(d_1\ket{0}+d_2\ket{1})_B\ket{1}_C$. If we get $\ket{\phi_{b2}}$, Alice perform a unitary transformation
\begin{equation}
    U_A =
    \lk
    \begin{array}{cc}
        1 & 0\\
        0 & -1
    \end{array}
    \rk,
    \qquad
\end{equation}
then we get $\ket{\phi_{b1}}$, too. So, with probability 1 we get $\ket{\phi_{b1}}$.

Thirdly, Charlie performs a POVM
\begin{equation}
    M_1 = \frac{1}{\sqrt{2}}
    \lk
    \begin{array}{cc}
        1 & e_1\\
        0 & e_2
    \end{array}
    \rk,
    \qquad
    M_2 = \frac{1}{\sqrt{2}}
    \lk
    \begin{array}{cc}
        1 & -e_1\\
        0 & -e_2
    \end{array}
    \rk,
\end{equation}
Then with probability $\frac{1}{2}$ we get $\ket{\phi_{c_1}}= \alpha\ket{000} + \beta\ket{1}_A(d_1\ket{0}+d_2\ket{1})_B(e_1\ket{0}+e_2\ket{1})_C$, which is exactly the $\ket\phi$ we want to get, and with probability $\frac{1}{2}$ we can get $\ket{\phi_{c_2}}= \alpha\ket{000} - \beta\ket{1}_A(d_1\ket{0}+d_2\ket{1})_B(e_1\ket{0}+e_2\ket{1})_C$. If we get $\ket{\phi_{c_2}}$, again, Alice can perform a unitary transformation
\begin{equation}
    U_A =
    \lk
    \begin{array}{cc}
        1 & 0\\
        0 & -1
    \end{array}
    \rk,
    \qquad
\end{equation}
 to get $\ket{\phi_{c1}}$, too. So with probability 1 we can get $\ket{\phi_{c1}}$. Then we can get $\ket\Psi$= $\alpha \ket{a_1b_1c_1} +\beta \ket{a_2b_2c_2}$ with certainty. \QED

Then we show the first three steps can be done with probability 1. For the last step, we have the following lemma.

\begin{lemma}
For $\ket\Psi$= $\gamma(\alpha \ket{a_1b_1c_1} +\beta \ket{a_2b_2c_2})$,
where $\alpha^2+\beta^2=1$ and $\gamma$ is a normalization factor, if
$<a_1|a_2>$=$\lambda_a$,
$<b_1|b_2>$=$\lambda_b$,$<c_1|c_2>$=$\lambda_c$, then there exists an LOCC
transformation protocol from GHZ state to $\ket\psi$ such that the
probability of success is at least
$\frac{1+2\alpha\beta\lambda_a\lambda_b\lambda_c}{1+\lambda_m}$,
where $\lambda_m$= $min(\lambda_a,\lambda_b,\lambda_c)$.
\end{lemma}

Proof: Firstly, we have $\ket\psi= \gamma(\alpha\ket{000} +
\beta(\lambda_a\ket{0}+\sqrt{1-\lambda^2_a}\ket{1})_A(\lambda_b\ket{0}+\sqrt
{1-\lambda^2_b}\ket{1})_B(\lambda_c\ket{0}+\sqrt{1-\lambda^2_c}\ket{1})_C)$,
where $\gamma=\frac{1}{\sqrt{1+2\alpha\beta\lambda_a\lambda_b\lambda_c}}$.
From theorem 1, we transform GHZ state to $\ket\xi=
\alpha\ket{000} +$
$\beta\ket{1}_A(\lambda_b\ket{0}$ $+\sqrt{1-\lambda^2_b}\ket{1})_B$
$(\lambda_c\ket{0} + \sqrt{1-\lambda^2_c}\ket{1})_C$,
with probability 1. Then from $\ket\xi$, Alice can do a POVM
\begin{eqnarray}
    &M_1 = \frac{1}{\sqrt{1+\lambda_a}}
    \lk
    \begin{array}{cc}
        1 & \lambda_a\\
        0 & \sqrt{1-\lambda^2_a}
    \end{array}
    \rk,
    \qquad\\
    &M_2 = \frac{\sqrt{\lambda_a}}{\sqrt{1+\lambda_a}}
    \lk
    \begin{array}{cc}
        1 & -1\\
        0 & 0
    \end{array}
    \rk,
\end{eqnarray}
So we have probability
$\frac{1+2\alpha\beta\lambda_a\lambda_b\lambda_c}{1+\lambda_a}$ to
get $\ket\psi$, and the other branch will give a state in which the
rank of $\rho_a$ is 1 so that the probability to get $\ket\psi$ from
it is zero. Then the total probability is
$\frac{1+2\alpha\beta\lambda_a\lambda_b\lambda_c}{1+\lambda_a}$.
However, we can do a permutation of A, B and C so that the
probability can also be
$\frac{1+2\alpha\beta\lambda_a\lambda_b\lambda_c}{1+\lambda_b}$ and
$\frac{1+2\alpha\beta\lambda_a\lambda_b\lambda_c}{1+\lambda_c}$. And
the maximum probability corresponds to
$min(\lambda_a,\lambda_b,\lambda_c)$. \QED

\begin{example}
Again take the transformation from $\ket{GHZ}$ to
$\ket{\phi} = \gamma(\ket{000} + \ket{aaa})$, where $\ket{a}=
c\ket{0}+ \sqrt{1-c^2}\ket{1}$ and $c \in (0,1]$ as an example, using the
protocol we provide, we can get a successful probability
$\frac{1+c^3}{1+c}$, let c=0.5, we have $p_s = 0.7500$. In comparison, the
upper bound we get in section 3 is 0.9604. There is still a gap between
these two values. How to reduce it is still an open problem.
\end{example}

Now we will generalize the result of lemma 5 higher dimensions and more parties. Suppose we are concerned with the transformation from the GHZ-state
generalized to n parties and m dimensions, $\ket{GHZmn}=
\frac{1}{\sqrt{m}} \sum^m_{i=1}\ket{i}_1\ket{i}_2...\ket{i}_n$ to
$\ket\psi$=
$\gamma(\sum^m_{i=1}\alpha_i\ket{k_{1_i}k_{2_i}...k_{n_i}})$. The
basic idea of our protocol can be divided into three steps: Firstly,
we would like to transform $\ket{GHZmn}$ into $\ket\Psi =
\sum^m_{i=1} \alpha_i \ket{i}\ket{i}\ket{i}$ which is called
GHZ-like (or Schmidt decomposable) state. Secondly, we transform
$\ket\Psi$ into $\ket{\psi_n}$ =
$\sum^m_{i=1}\alpha_i\ket{i_1k_{2_i}k_{i_3}...k_{i_n}}$. We will
show these two steps can be done with probability 1. Finally, if for
at least m-1 terms of $\ket\psi$, there can be at least one party
with a state that is orthogonal to this party's state in every other
term, our protocol can transform $\ket{GHZmn}$ into $\ket\psi$ with
probability 1. In other cases, our protocol can be done with a
probability higher than what have been known before.

The second and third step for the case when for at least m-1 terms
of $\ket\psi$, there can be at least one party with a state that is
orthogonal to this party's state in every other term are
incorporated in the following theorem.

\begin{theorem}
\label{lem:ghz_n_orth}
The GHZ-state generalized to n parties and m dimensions, $\ket{GHZmn}=
\frac{1}{\sqrt{m}} \sum^m_{i=1}\ket{i}_1\ket{i}_2...\ket{i}_n$, can be
transformed to $\ket\psi$=
$\sum^m_{i=1}\alpha_i\ket{k_{1_i}k_{2_i}...k_{n_i}}$ with probability 1, if
$\exists$ p, $\forall$ i$\neq$ p, $\exists$ j with
$\braket{k_{j_i}|k_{j_l}}=0$ for $\forall$ $l\neq i$. This means for at
least m-1 terms, there has to be at least one party with a state that is
orthogonal to this party's state in every other term.
\end{theorem}

Proof: The basic idea is we at first make the coefficient of each term equal
to the corresponding terms of the destination state. Then let party 2
perform a POVM which transforms the state into many states in such a form:
for each term, the party 2 part of the term is the same with the destination
state, but the term's coefficient maybe of the same or the opposite sign of the
destination state. Then by introducing a unitary transformation on the party
1, we can make all the coefficients the same with the destination state.
Keep doing this for party 3, $\cdot$, n. Finally, do a similar POVM on party
1, we can get many states in such a form: the corresponding terms are the
same, but the coefficients may be of the same or the opposite sign. Then we can use
unitary transformation to transform all the states into the destination
state. Exact process is in the following:

Firstly, using the result of \cite{Xin2007} to get $\ket{GHZmn}=
\gamma(\sum^m_{i=1}\alpha_i\ket{i}_1\ket{i}_2...\ket{i}_n)$ where
$\gamma$ is a normalization factor. But then, the POVMs should be
modified. We call the parties 1,2, $\cdots$, n party 1, party 2, and
so on. Take Party 2 as an example, suppose with a unitary transformation,
$\ket{k_{2_i}}=\sum^i_{j=0}a_{2_{ij}}\ket{j}_2$, in which
$a_{2_{00}}=1$, then party 2 can operate a POVM
\begin{gather}
\begin{split}
    &M_0 = \frac{1}{\sqrt{2^{m-1}}}
    \lk
    \begin{array}{ccccc}
        1 & a_{2_{10}} &\cdots & a_{2_{(m-1)0}} \\
        0 & a_{2_{11}} &\cdots & a_{2_{(m-1)1}} \\
        \vdots & \vdots & \vdots & \vdots \\
        0 & 0       &\cdots & a_{2_{(m-1)(m-1)}}
    \end{array}
    \rk,
    \qquad \\
    &M_1= \frac{1}{\sqrt{2^{m-1}}}
    \lk
    \begin{array}{ccccc}
        1 & a_{2_{10}} &\cdots & -a_{2_{(m-1)0}} \\
        0 & a_{2_{11}} &\cdots & -a_{2_{(m-1)1}} \\
        \vdots & \vdots & \vdots & \vdots \\
        0 & 0       &\cdots & -a_{2_{(m-1)(m-1)}}
    \end{array}
    \rk,
    \qquad \\
    &\vdots \\
    &M_{2^{m-1}-1}= \frac{1}{\sqrt{2^{m-1}}}
    \lk
    \begin{array}{ccccc}
        1 & -a_{2_{10}} &\cdots  & -a_{2_{(m-1)0}} \\
        0 & -a_{2_{11}} &\cdots  & -a_{2_{(m-1)1}} \\
        \vdots & \vdots & \vdots & \vdots \\
        0 & 0       &\cdots  & -a_{2_{(m-1)(m-1)}}
    \end{array}
    \rk
    \end{split}
\end{gather}
Then we can get a state unitary equivalent with $\ket{\psi_2}$=
$\sum^m_{i=1}\alpha_i\ket{i_1k_{i_2}i_3...i_n}$ with probability
$\frac{1}{2^{m-1}}$ and with the same probability we get other states which
are different with $\ket{\psi_2}$ just because some terms have an opposite
sign than the corresponding terms in $\ket{\psi_2}$.

And we do a unitary transformation on party 1 to transform all
branches into $\ket{\psi_2}$. For party 3,4, $\cdots$ n, we can use
similar method, so at last we can get a state unitary equivalent
with $\ket{\psi_n}$=
$\sum^m_{i=1}\alpha_i\ket{i_1k_{2_i}k_{i_3}...k_{i_n}}$. Then we
finish the second step with probability 1.

After that, party 1 can perform a similar POVM and with probability
$\frac{1}{2^{m-1}}$ we get $\ket\psi$ which we want and with the
same probability we get other states which are different with
$\ket\psi$ just because some terms have an opposite sign than the
corresponding terms in $\ket\psi$.

However, if for at least m-1 terms, there is at least one party with
a state, which we call $\ket{k_{i_j}}$, that is orthogonal to this
party's state in every other term, we can introduce a minus sign for
this term by a unitary transformation of party j which transforms
$\ket{k_{i_j}}$ to $-\ket{k_{i_j}}$ and do nothing to all the other
states orthogonal to $\ket{k_{i_j}}$. Thus we can introduce a minus
sign for these m-1 terms just by unitary transformation. For the
only one term which does not have this property (if it exists), we
can introduce a minus sign for every other term and then multiply -1
for the whole wave function. Then, we can get $\ket\psi$ with
probability 1. \QED

\begin{remark}
The condition we require in
Lemma~\ref{lem:ghz_n_orth} is different
from each term is orthogonal to other ones. In fact, it is a
stronger requirement than orthogonality. See the example below: for a
state $\ket\phi = \frac{1}{\sqrt{3}}[\ket{000} +
\ket{1}(\ket{0}+\ket{1})\ket{0} + (\ket{0}+\ket{1})\ket{0}\ket{1}]$,
it is easy to check each term is orthogonal to another in this
state. But we do not know how to introduce a minus sign for any term
because the condition in Lemma 1 is not satisfied.
\end{remark}

There is an open question: Can this condition in Lemma 2 be "if and
only if" in higher dimensions? To make it also "only if", there are
two problems: 1. Is the form we write the state still unique in
higher dimensions? We know, for a 2-term tripartite state in which
each party has rank 2, if we write it in the form $\ket\psi=
\gamma(\alpha\ket{000} +
\beta(\lambda_a\ket{0}+\sqrt{1-\lambda^2_a}\ket{1})_A(\lambda_b\ket{0}+\sqrt
{1-\lambda^2_b}\ket{1})_B(\lambda_c\ket{0}+\sqrt{1-\lambda^2_c}\ket{1})_C)$,
the result should be unique. It is also true for a 3-term tripartite
state (the W-class state) \cite{Ac'in2000}, but the similar result for the higher dimension conditions
have not been proved. 2. Can a state in which all terms are
orthogonal to each other be transformed from GHZ like state with
probability 1? We know, our protocol can only work for a stronger
requirement. However, for states with orthogonal terms, the inner
product is also zero, so how can we prove the probability can not be
one in this case is another problem.

\begin{corollary}
All tripartite pure three qubit states can be transformed from 3-term GHZ
state
$\frac{1}{\sqrt{3}}(\ket{000}+\ket{111}+\ket{222})$ with probability
1.
\end{corollary}
Proof: From the paper \cite{Ac'in2000}, we know any tripartite
pure state can be written as
\begin{gather}
\begin{split}
\ket{\Phi}= &\lambda_0\ket{000} + \lambda_1e^{i\phi}\ket{100} + \lambda_2|101>\\
            &+\lambda_3\ket{110} +\lambda_4\ket{111},\\
            &\lambda_i\geq0, 0\leq\phi\leq\pi,
\mu_i\equiv\lambda^2_i,\sum{\mu_i}
= 1
\end{split}
\end{gather}

And shown in \cite{Ac'in2000}, if Charlie introduces a unitary
transformation
\begin{equation}
    U = \frac{1}{\sqrt{\mu_1+\mu_2}}
    \lk
    \begin{array}{cc}
        \lambda_1 e^{-i\phi} & \lambda_2 e^{-i\phi}\\
        \lambda_2 & -\lambda_1
    \end{array}
    \rk,
\end{equation}

we can get
\begin{gather}
\begin{split}
\ket{\Psi}= &
\frac{1}{\sqrt{\mu_1+\mu_2}}[e^{-i\phi}\lambda_0\lambda_1\ket{000} +
e^{-i\phi}\lambda_0\lambda_2\ket{001}\\
            & +(\lambda_1^2+\lambda_2^2)\ket{100} + (\lambda_1\lambda_3 + \lambda_2\lambda_4)\ket{110}\\
            & +
(\lambda_2\lambda_3 - \lambda_1\lambda_4)\ket{111}]
\end{split}
\end{gather}
which is unitary equivalent with the state we want.

If we combine the first and second term into a term and do the same
for third and fifth term we can get $|\Psi>=
\ket{00}(a\ket{0}+b\ket{1}) + d\ket{100} +
\ket{11}(c\ket{0}+e\ket{1})$, easy to see, if we consider it as a 3-term
state, it satisfies the condition we required for Lemma 2, (Alice in
first term and Bob in third term), so we can transform it from
3-term GHZ generalized state with probability 1. \QED

% \begin{corollary}
% W class state $(\sqrt{c}\ket{1}+ \sqrt{d}\ket{0})\ket{00}+
% \ket{0}(\sqrt{a}\ket{01}+\sqrt{b}\ket{10})$, where c+d+a+b=1, can be
% transformed from 3-term GHZ state
% $\frac{1}{\sqrt{3}}(\ket{000}+\ket{111}+\ket{222})$ with probability
% 1.
% \end{corollary}
% Proof: As W class satisfies the condition in
% Lemma~\ref{lem:ghz_n_orth}, it
% can be transformed from a 3-term GHZ state
% $\frac{1}{\sqrt{3}}(\ket{000}+\ket{111}+\ket{222})$ with probability
% 1.

\begin{theorem}
For a general $\ket\psi$=
$\gamma(\sum^m_{i=1}\alpha_i\ket{k_{1_i}k_{2_i}...k_{n_i}}),$ where
$\sum^m_{i=1}\alpha^2_i=1$ and $\gamma$ is the normalization factor.
There exists an LOCC transformation protocol from the GHZ-state
generalized to n parties and m dimensions, $\ket{GHZmn}=
\frac{1}{\sqrt{m}} \sum^m_{i=1}\ket{i}_1\ket{i}_2...\ket{i}_n$ to
$\ket\psi$ such that the probability of success is at least
$max(\frac{1}{\gamma||A_i||^2})$, where
\begin{equation}
 A_i = \frac{1}{\sqrt{2^{m-1}}}
    \lk
    \begin{array}{ccccc}
        1 & a_{i_{10}} &\cdots & a_{i_{(m-2)0}} & a_{i_{(m-1)0}} \\
        0 & a_{i_{11}} &\cdots & a_{i_{(m-2)1}} & a_{i_{(m-1)1}} \\
        \vdots & \vdots & \vdots & \vdots & \vdots \\
        0 & 0       &\cdots & 0 & a_{i_{(m-1)(m-1)}}
    \end{array}
    \rk
\end{equation}
\end{theorem}

Proof: To generalize theorem 2 to general m-term n-party states, we
can firstly get $\ket{\psi_n}$=
$\sum^m_{i=1}\alpha_i\ket{i_1k_{i_2}k_{i_3}...k_{i_n}}$ with
certainty. And Alice perform a POVM
$\{\frac{A_1}{||A_1||},\sqrt{1-\frac{A_1^+A_1}{||A_1||^2}}\}$, where
\begin{equation}
    A_1 = \frac{1}{\sqrt{2^{m-1}}}
    \lk
    \begin{array}{ccccc}
          1 & a_{1_{10}} &\cdots & a_{1_{(m-2)0}} & a_{1_{(m-1)0}} \\
        0 & a_{1_{11}} &\cdots & a_{1_{(m-2)1}} & a_{1_{(m-1)1}} \\
        \vdots & \vdots & \vdots & \vdots & \vdots \\
        0 & 0       &\cdots & 0 & a_{1_{(m-1)(m-1)}}
    \end{array}
    \rk
\end{equation}

After calculation we can find the successful probability is
$\frac{1}{\gamma||A_1||^2}$. Similarly, we can choose other party to
finish the final step and find the best one which give the maximum
transformation probability.

\section{Summary and Concluding Remarks}
\label{sec:summary}

We derive upper bound and lower bound for the supremum transformation
probability from GHZ state to GHZ-class state. In the derivation of the upper
bounds, we consider the action of the LOCC protocol on a different input
state, namely $1/\sqrt{2} [\ket{000} - \ket{111}]$, and demand that
the probability of an outcome remains bounded by 1. By considering the constraints of the interference term and 3-tangle, we find an upper bound for more general cases. For the lower bound, we construct a new transformation protocol: the four-step method to do the transformation. Before that, there was no
nontrivial upper bound known for this transformation. Based on the previous
results of weak measurement, we construct a "stop and reconstruct" method
which may be very useful in the analyzation of the LOCC transformation
protocols. The lower bound is generalized into higher dimension. During the
discussion of lower bound, we find all tripartite pure 3-qubit states can be transformed
from
$\ket{GHZ_{23}} = \frac{1}{\sqrt{3}}(\ket{000} + \ket{111} + \ket{222})$
with probability 1. This is a new result.

There are still open questions and possible future generalization of the
result we have, which is mentioned during the above discussion. To
summarize, firstly, there is still a gap between the upper bound and lower
bound we get. How to further decrease the gap and finally find the optimum
transformation protocol are still open questions. Secondly, we want to generalize the
upper bound we get to higher dimension. To do this, we need firstly make
sure the GHZ-class state generalized into the higher dimension is still
unique so that we can still talk about the inner product. We also need to find the corresponding entanglement monotone for higher dimension case. Finally, the result of Corollary 1 is very surprising, one
question is whether it can be generalized into higher dimension case. To do
this, we need to analyze the Generalized Schmidt Decomposition in higher
dimension \cite{Tamaryan2008} or other forms to express the state in higher
dimensions \cite{Verstraete2002}. 

\section{Appendix: Proof of Theorem 5}
\label{sec: appendix}

Proof: From the formula of the two quantity:
\begin{eqnarray}
\label{eq:interferencetermdefi}
I =&
\frac{2c_{\alpha}c_{\beta}c_{\gamma}s_{\delta}c_{\delta}c_{\varphi}}{1+2c_{
\alpha}c_{\beta}c_{\gamma}s_{\delta}c_{\delta}c_{\varphi}}\\
I^2 =&
\frac{4c^2_{\alpha}c^2_{\beta}c^2_{\gamma}s^2_{\delta}c^2_{\delta}c^2_{
\varphi}}{(1+2c_{\alpha}c_{\beta}c_{\gamma}s_{\delta}c_{\delta}c_{\varphi}
)^2}\\
\tau_{ABC} =&
\frac{4s^2_{\alpha}s^2_{\beta}s^2_{\gamma}s^2_{\delta}c^2_{\delta}}{(1+2c_{
\alpha}c_{\beta}c_{\gamma}s_{\delta}c_{\delta}c_{\varphi})^2}\\
\end{eqnarray}

We have
\begin{eqnarray}
\tau_{ABC} =& I^2
\frac{s^2_{\alpha}s^2_{\beta}s^2_{\gamma}}{c^2_{\alpha}c^2_{
\beta}c^2_{\gamma}c^2_{\varphi}}\notag\\
           =& I^2
\frac{(1-c^2_{\alpha})(1-c^2_{\beta})(1-c^2_{\gamma})}{c^2_{
\alpha}c^2_{\beta}c^2_{\gamma}c^2_{\varphi}}
\end{eqnarray}

We consider the condition when $I > 0$ at first.

Firstly, we consider the condition when $s_{\delta} = c_{\delta} = \frac{\sqrt{2}}{2}, c_{\varphi} =1$. In this case, we have $c_{\alpha}c_{\beta}c_{\gamma} = \frac{I}{1-I}$. let $\frac{I}{1-I} = a^3$ where $a = (\frac{I}{1 - I})^{\frac{1}{3}}$. Then we have

\begin{eqnarray}
\tau_{ABC} =& I^2 \frac{(1-c^2_{\alpha})(1-c^2_{\beta})(1-c^2_{\gamma})}{c^2_{
\alpha}c^2_{\beta}c^2_{\gamma}}\notag\\
           =& I^2 \frac{(1-c^2_{\alpha})(1-c^2_{\beta})(1-\frac{a^6}{c^2_{
\alpha}c^2_{\beta}})}{a^6}
\end{eqnarray}

Take partial derivation of $c_{\alpha}$ and $c_{\beta}$ we can find this expression reaches its maximum value when $c_{\alpha} = c_{\beta} = c_{\gamma} = a$ and the corresponding maximum value of 3-tangle is $\tau_{{ABC}_0} = ((1-I)^{\frac{2}{3}} - I^{\frac{2}{3}})^3 = \frac{(1 - a^2)^3}{(1 + a^3)^2}$.

Now we will show in other cases when $s_{\delta} \neq c_{\delta} or c_{\varphi} < 1$, we can only get a 3-tangle smaller than $\tau_{{ABC}_0}$.

If $s_{\delta} \neq c_{\delta}$, we will have $s_{\delta}c_{\delta} < \frac{1}{2}$, then from the expression of I we can find $c_{\alpha}c_{\beta}c_{\gamma}c_{\varphi} > \frac{I}{1-I} = a^3$, then as $c_{\varphi} \leq 1$, we also have $c_{\alpha}c_{\beta}c_{\gamma} = b^3 > a^3$. And also take the partial derivation of $(1-c^2_{\alpha})(1-c^2_{\beta})(1-c^2_{\gamma})$ we can find its maximum value is $(1 - b^2)^3 < (1 - a^2)^3$. Finally we have
\begin{eqnarray}
\tau_{ABC} =& I^2 \frac{(1-c^2_{\alpha})(1-c^2_{\beta})(1-c^2_{\gamma})}{c^2_{
\alpha}c^2_{\beta}c^2_{\gamma}c^2_{\varphi}}\notag\\
           <& I^2 \frac{(1-a^2)^3}{a^6} = \frac{(1 - a^2)^3}{(1 + a^3)^2} = \tau_{{ABC}_0}
\end{eqnarray}

That is to say, when $s_{\delta} \neq c_{\delta}$, $\tau_{ABC}$ is always smaller than $\tau_{{ABC}_0}$. Now let us consider the case when $s_{\delta} = c_{\delta} = \frac{\sqrt{2}}{2}$, but $c_{\varphi} < 1$.

Then again we have $c_{\alpha}c_{\beta}c_{\gamma}c_{\varphi} = \frac{I}{1-I} = a^3$. But as $c_{\varphi} < 1$, we still have $c_{\alpha}c_{\beta}c_{\gamma} = d^3 > a^3$. And also take the partial derivation of $(1-c^2_{\alpha})(1-c^2_{\beta})(1-c^2_{\gamma})$ we can find its maximum value is $(1 - d^2)^3 < (1 - a^2)^3$. So we have

\begin{eqnarray}
\tau_{ABC} =& I^2 \frac{(1-c^2_{\alpha})(1-c^2_{\beta})(1-c^2_{\gamma})}{c^2_{
\alpha}c^2_{\beta}c^2_{\gamma}c^2_{\varphi}}= I^2 \frac{(1-d^2)^3}{a^6}\notag\\
           <& I^2 \frac{(1-a^2)^3}{a^6}
           = \frac{(1 - a^2)^3}{(1 + a^3)^2} = \tau_{{ABC}_0}
\end{eqnarray}

Then we show, for the interference term $I > 0$, we have
\begin{equation}
max(\tau_{ABC}(\phi | I(\phi) = I > 0)) =
\frac{(1-a^2)^3}{(1+a^3)^2}
\end{equation}

When $I \leq 0$, the discussion is almost the same. Except that, we need to consider the condition $s_{\delta} = c_{\delta} = \frac{\sqrt{2}}{2}, c_{\varphi} =-1$ first and find $c_{\alpha}c_{\beta}c_{\gamma} = -\frac{I}{1-I} = a'^3 > 0$. Then easy to find the corresponding maximum value is $\frac{(1-a'^2)^3}{(1-a'^3)^2}$. And use the same tricks one can show it is the maximum value of the 3-tangle.

One thing to notice is that, the expression of a' and a is just opposite to each other. So if we let $a = \frac{I}{1-I} = -a'$ when $I \leq 0$, we will get

\begin{equation}
max(\tau_{ABC}(\phi | I(\phi) = I \leq 0)) = \frac{(1-a'^2)^3}{(1-a'^3)^2} = \frac{(1-a^2)^3}{(1+a^3)^2}
\end{equation}

Then in all we have
\begin{equation}
max(\tau_{ABC}(\phi | I(\phi) = I) = \frac{(1-a'^2)^3}{(1-a'^3)^2} = \frac{(1-a^2)^3}{(1+a^3)^2}
\end{equation}

\QED.


\begin{thebibliography}{10}

\bibitem{Bennett1996}
C.~H. Bennett, H.~J. Bernsein, S. Popescu, and B. Schumacher, Phys. Rev. A
  {\bf 53},  2046  (1996).

\bibitem{Lo2001}
H.-K. Lo and S. Popescu, Phys. Rev. A {\bf 63},  022301  (2001).

\bibitem{Nielsen1999}
M.~A. Nielsen, Physical Review Letters {\bf 83},  436  (1999).

\bibitem{Vidal1999}
G. Vidal, Phys. Rev. Lett. {\bf 83},  1046  (1999).

\bibitem{Dur2000}
W. D\"ur, G. Vidal, and J.~I. Cirac, Phys. Rev. A {\bf 62},  062314  (2000).

\bibitem{Verstraete2002}
F. Verstraete, J. Dehaene, B. {De Moor}, and H. Verschelde, Phys. Rev. A {\bf
  65},  052112  (2002).

\bibitem{Ac'in2000a}
A. Ac\'in, E. Jan\'e, W. D\"ur, and G. Vidal, Phys. Rev. Lett. {\bf 85},  4811
  (2000).

\bibitem{Turgut2009}
S. Turgut, Y. Gul, and N.~K. Pak, arXiv:0907.3960v2  (2009).

\bibitem{Eisert2001}
J. Eisert and H.~J. Briegel, Phys. Rev. A {\bf 64},  022306  (2001).

\bibitem{Ac'in2000}
A. Ac\'in {\it et~al.}, Phys. Rev. Lett. {\bf 85},  1560  (2000).

\bibitem{Coffman2000}
V. Coffman, J. Kundu, and W.~K. Wootters, Phys. Rev. A {\bf 61},  052306
  (2000).

\bibitem{Oreshkov2005}
O. Oreshkov and T.~A. Brun, Physical Review Letters {\bf 95},  110409  (2005).

\bibitem{Andersson2007}
E. Andersson and D.~K.~L. Oi, arXiv:0712.2665  (2007).

\bibitem{Chitambar2008}
E. Chitambar, R. Duan, and Y. Shi, Phys. Rev. Lett. {\bf 101},  140502  (2008).

\bibitem{Xin2007}
Y. Xin and R. Duan, arxiv:0707.1947  (2007).

\bibitem{Tamaryan2008}
L. Tamaryan, D. Park, and S. Tamaryan,   (2008).

\end{thebibliography}
\end{document}